\documentclass[12pt]{iopart}

\usepackage{graphicx}
\usepackage{siunitx}
\usepackage{enumerate}

\begin{document}

\title[Biophysical constraints and the evolution of phenotypic fluctuations]{Biophysical constraints determine the selection of phenotypic fluctuations during directed evolution}
%Genetic and environmental structure determine the evolution of phenotypic fluctuations
\author{Hong-Yan Shih, Harry Mickalide, David T. Fraebel, Nigel Goldenfeld$^*$, Seppe Kuehn$^*$}

\address{Department of Physics and Center for the Physics of Living Cells, Loomis Laboratory of Physics,
University of Illinois at Urbana-Champaign, 1110 West Green St.,
Urbana, IL 61801, USA}
%\ead{submissions@iop.org}
\vspace{10pt}
\begin{indented}
\item[$^*$ To whom correspondence should be addressed:]%February 2014
\item[SK (seppe@illinois.edu); NG (nigel@uiuc.edu)]
\end{indented}

\begin{abstract}
Phenotypes of individuals in a population of organisms are not fixed.  Phenotypic fluctuations, which describe temporal variation of the phenotype of an individual or individual-to-individual variation across a population, are present in populations from microbes to higher animals.  Phenotypic fluctuations can provide a basis for adaptation and be the target of selection.  Here we present a theoretical and experimental investigation of the fate of phenotypic fluctuations in directed evolution experiments where phenotypes are subject to constraints.  We show that selecting bacterial populations for fast migration through a porous environment drives a reduction in cell-to-cell variation across the population.  Using sequencing and genetic engineering we study the genetic basis for this reduction in phenotypic fluctuations.  We study the generality of this reduction by developing a simple, abstracted, numerical simulation model of the evolution of phenotypic fluctuations subject to constraints.  Using this model we find that strong and weak selection generally lead respectively to increasing or decreasing cell-to-cell variation as a result of a bound on the selected phenotype under a wide range of parameters.  However, other behaviors are also possible, and we describe the outcome of selection simulations for different model parameters and suggest future experiments.  We analyze the mechanism of the observed reduction of phenotypic fluctuations in our experimental system, discuss the relevance of our abstract model to the experiment and explore its broader implications for evolution.
\end{abstract}


\vspace{2pc}

\submitto{\PB}

\maketitle


\section{Introduction}

Natural selection acts at the level of the phenotype.  Unlike genomes,
phenotypes can be highly variable over the lifetime of a single
organism or heterogeneous across a genetically identical population.
Given the central role of the phenotype in selection, phenotypic
fluctuations are believed to play an important role in evolution.

Therefore, understanding the
evolutionary origins and impacts of phenotypic fluctuations will be
central to any quantitative theory of evolution.  Environmental factors provide selection pressure that prefers certain phenotypes, through which the mutant genotypes that represent similar phenotypes can be selected.  Phenotypic fluctuations can arise by stochastic variation in gene
expression \cite{Elowitz:2002hb}, which can be associated with physiological responses to environmental variation
(plasticity)\cite{Lehner:2010ei}.  In bacteria, non-genetic phenotypic variability in a population is critical for survival in the presence of antibiotics\cite{Balaban:2004bq}.  Further, non-genetic variation is present in bacterial swimming behavior\cite{Spudich:1976wv} and is thought to be adaptive\cite{Frankel:2014fb}.   

The role of phenotypic fluctuations in evolution, and how genetic
variation alters phenotypic fluctuations, has been the subject of
theoretical and experimental investigations since Baldwin \cite{Baldwin:1896dt}.  Waddington presented
compelling arguments for the role of phenotypic plasticity in
facilitating evolution through genetic assimilation \cite{Waddington:1953vf}, and
conceptual models of this effect abound \cite{WestEberhard:1989wq}.  Notably,
Kaneko formulated a phenomenological model based on the
fluctuation-dissipation theorem, which postulates that phenotypes
exhibiting larger fluctuations should evolve more rapidly under
selection \cite{Sato:2003uu}.  The theory was tested in a directed evolution experiment by constructing a diverse population of green fluorescent protein (GFP) expressing \textit{Escherichia coli} mutants synthetically and then selecting for higher levels of GFP fluorescence.  The study showed that directed selection for increasing mean fluorescence resulted in reduced cell-to-cell variability in fluorescence intensity \cite{Sato:2003uu}.    
Conversely, a subsequent series of experimental studies showed that strong selection on the phenotype led to an increase in phenotypic fluctuations \cite{kaneko2009selection}.  The interpretation of this experiment is complicated, however, because there were only a few clones in the system, and the population seemed to split into two types of mutant distinguished by the variance in their phenotype fluctuations \cite{kaneko2009selection}.  Similarly, in directed evolution experiments of cell size in \textit{E. coli} a decrease of cell-to-cell variation in size was reported for weak selection whereas little change in cell size fluctuation was observed under strong selection \cite{yomo2014}.  

Phenotypes arise from genotypes through the processes of transcription and translation.  Therefore, any generic features of the evolution of phenotypic fluctuations might be illuminated by considering universal aspects of gene expression.  Protein copy number distributions have been measured in a variety of microbial species, for example in cultured populations of bacteria \cite{kaneko2005,Salman:2012bb,cohen2015} and yeast \cite{Salman:2012bb,cohen2015} and in single-cells \cite{taniguchi2010,salman2015single,salman2016}.  These studies show that the probability density of protein copy number across a population is consistently non-Gaussian and highly skewed, and reportedly well fit by gamma \cite{taniguchi2010}, extreme value (Fisher-Tippett-Gumbel \cite{cohen2015} or Frechet \cite{Salman:2012bb}) or log-normal \cite{kaneko2005,salman2016} distributions, all of which are similar in shape.  Regardless of the precise form of the distribution reported, one trend is clear: the standard deviation $\sigma$ is a monotonically increasing function of the mean, and the distributions can be collapsed onto a single universal curve using reduced coordinates ($n-\langle n\rangle)/\sigma$ \cite{Salman:2012bb,salman2016}.  If a phenotype can be associated with a particular dominant protein, then as the phenotype and hence the protein copy number is increased during a directed evolution experiment, one might naively expect the phenotypic variation to increase as well, a result that is not generically found to be true.  In reality, the relationship between protein copy number and phenotype is more complex, reflecting regulation, inhibition, and feedback.  Therefore, the precise relationship between protein copy number and phenotype remains unclear, with little likelihood of a universal connection, even if the global statistics exhibit universal functional forms. 

Direct empirical evidence for the relationship between
phenotypic fluctuations and long-term evolution remains limited.  Notable
exceptions include retrospective studies of hemoglobin binding affinity
across mammals\cite{Milo:2007wm}, but even this study does not make direct measurements of phenotypic fluctuations in time or across individuals.  While experimental evolution has revealed
striking examples of phenotypic evolution\cite{Barrick:2009in,Rainey:1998fx,Bachmann:2013dm,Fraebel:2017wx},
quantitative measurements of phenotypic fluctuations in many of these
experiments have not been made.  As a result, conceptual or
quantitative models of the evolution of phenotypic fluctuations remain
untested. 

Here we present a joint theoretical and experimental
investigation of how phenotypic fluctuations evolve under selection.  We
use high-throughput phenotyping to show that the phenotypic
variation in the population declines when bacteria are selected
for faster migration through a porous environment\cite{Fraebel:2017wx}.  We then present a simple model of directed evolution which allows us to interrogate
how selection strength and mutations result in the evolution of phenotypic
fluctuations.  We show that, depending on the strength of selection, phenotypic fluctuations can decline when phenotypes are subjected to constraints even when there is no mechanistic link between the mean trait value and phenotypic fluctuations.  We discuss the relevance of this theoretical result to the experimentally observed reduction in phenotypic fluctuations.  Finally, we discuss the possible biological mechanisms underlying the experimentally observed reduction in phenotypic fluctuations in the context of our model.

%Whole genome sequencing of the strains which
%evolve reduced phenotypic variation in our experiment along with genetic engineering of individual mutations suggests
%potentially generic mechanisms for the evolution of phenotypic
%variability.  We postulate that specific genetic regulatory
%architectures are poised for the evolution of phenotypic variability. 

% MOVE ABSTRACT MODEL TO AFTER THE STOCHASTIC MODEL.

\section{Evolution of faster migration in \textit{E. coli}}
\begin{figure*}[ht!]
\begin{centering}
\includegraphics[scale = 0.35]{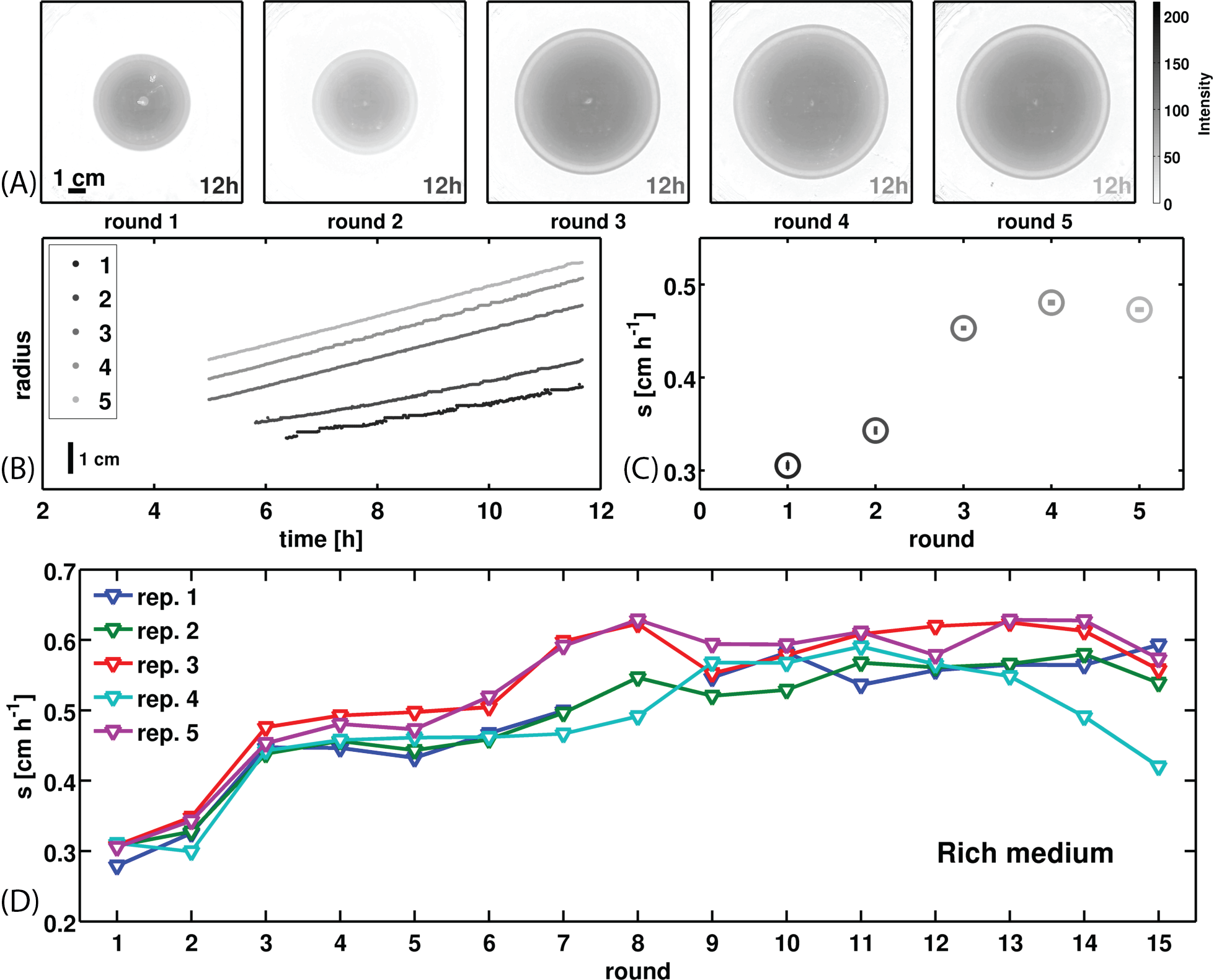}
\caption{\textbf{Selection for faster bacterial migration:} (A) Shows images of \textit{E. coli} colonies in low viscosity agar plates  after \num{12} hours of expansion.  After \num{12} hours a sample of the outer band of cells is taken and approximately \num{e6} cells are used to initiate another identical agar plate (second panel).  This process is repeated for \num{15} rounds of selection where a round consists of colony expansion in a single plate.  The color bar to the right applies to all panels, with darker gray indicating higher cell density.  Scale bar in left panel applies to all panels in (A). (B) The radius of each colony in (A) as a function of time, lighter shades of gray denote later rounds of selection and correspond to labels in (A).  Traces are offset vertically for clarity, note scale bar lower left.  (C) The rate of the linear portion of the colony expansion as a function of the round of selection for the plates shown in (A-B). (D) The evolutionary process outlined in (A-C) was carried out in five independent experiments.  Each line corresponds to an independent selection experiment.  Round \num{8} for replicate \num{1} is missing due to failure of the imaging device.  The data in panels (A-C) are from replicate \num{5}.  Errors in rate of expansion are smaller than the size of markers.  Data recapitulated from \cite{Fraebel:2017wx} \label{Figure1}}
\end{centering}
\end{figure*}

Growing populations of motile, chemotactic bacteria migrate outward when inoculated into a soft agar plate containing growth medium and a chemoattractant\cite{Wolfe:1989ue,Adler:1966wi}. As cells swim and divide in this porous environment local depletion of nutrients establishes a spatial nutrient gradient which drives chemotaxis through the three-dimensional agar matrix and subsequent nutrient consumption.  Microscopically, cells move through the porous environment by executing runs, at a speed $|v_r| \sim$ \SI{20}{\micro\meter\per\second} for a run duration $\tau_r\sim$ \SI{1}{\second}, and tumbles which rapidly reorient the cell in $\tau_t\sim$ \SI{0.1}{\second}. Tumbles are essential for avoiding obstacles in order to successfully navigate the soft agar\cite{Wolfe:1989ue}. The result is a macroscopic colony that expands radially through the bulk of the plate at a constant speed after an initial growth phase. We selected populations of \textit{E. coli} (MG1655-motile, Coli Genetic Stock Center, Yale University \#\num{6300}) for faster migration through soft agar by repeatedly allowing a colony to expand for a fixed interval, sampling a small population of cells from its outer edge and using a portion of this sample to inoculate a new plate while preserving the remainder cryogenically (Fig.~\ref{Figure1}). In rich medium conditions (LB, \num{0.3}\% w/v agar, \SI{30}{\degreeCelsius}), we sampled after \num{12} hours of expansion for a total of \num{15} rounds of selection. By performing time-lapse imaging on the expanding colonies, we found that the migration rate approximately doubled over the first five rounds of selection and continued to increase marginally in subsequent rounds. We found that this increase was reproducible across replicate experiments.

To understand the mechanism by which faster migration evolved, we performed single cell tracking on hundreds of individuals from the ancestral strain as well as from strains isolated after \num{5}, \num{10} and \num{15} rounds of selection. Individual cells were trapped in a circular microfluidic chamber in the same medium in which the selection was performed and recorded while swimming for \num{5} minutes per cell.  Swimming cells were imaged at \SI{30}{\hertz}, automated tracking routines constructed swimming trajectories from these movies and runs and tumbles were automatically identified as described previously\cite{Fraebel:2017wx}.  This measurement permitted us to capture the swimming behavior of hundreds of single bacterial cells in the absence of chemical gradients.  We found that the average run speed increased by approximately \SI{50}{\percent} during selection, while the duration of run and tumble events declined (Fig. \ref{fig_experiment_distribution}).  The maximum growth rate, which was measured in a separate experiment by monitoring the optical density of a well-stirred liquid culture declined over the course of selection.  The trade-off between swimming speed and growth rate is the subject of a separate study\cite{Fraebel:2017wx} and similar trade-offs have been observed elsewhere\cite{Yi:2016dp}.

\section{Phenotypic fluctuations decline with selection}

Phenotypic fluctuations have previously been characterized in several ways.  In some cases, fluctuations refer to the time-dependence of a specific phenotypic parameter during the lifetime of an individual\cite{Korobkova:2004vs}.   In other studies, fluctuations refer to cell-to-cell variation in time-averaged phenotypic measurements over a population\cite{Bai:2013eea,Jordan:2013hf,Spudich:1976wv,wagner1996}.  Here we use the latter approach, which is shown schematically in Fig. \ref{fig_illustration_distribution}.  Briefly, from run-tumble events performed by each individual cell we computed an average phenotype (run duration, tumble duration, and run speed) for each cell.  From these data we computed a distribution of average phenotypes across individuals in the population, and thus extracted the standard deviation over the population of a given phenotype.  This standard deviation directly measures cell-to-cell variation, as sketched in Fig. \ref{fig_illustration_distribution}.  

To define phenotypic fluctuations more explicitly consider a single \textit{E. coli} cell which exhibits a series of runs and tumbles.  Each run event is described by a run duration ($\tau_r$) and a run speed ($|v_r|$) and each tumble by a tumble duration ($\tau_t$) and an angular velocity ($\omega_t$).  Even in an unstimulated environment where no gradients are present, $\tau_r$ will vary between run events, and the distribution exhibited by individual $i$ is given by $P(\tau_r^{(i)})$.  Each run event  for this individual has a duration drawn from this distribution.  Similar distributions exist for $|v_r|$, $\tau_t$, and $\omega_t$, but $\omega_t$ is difficult to measure accurately for single cells, and we omit this parameter from consideration.  We consider the phenotype of a single cell to be the mean of these distributions.  Thus a complete description of unstimulated swimming behavior of a single cell is captured by the set of phenotypes $\chi^{(i)} \in \{ \langle \tau_r \rangle^{(i)}, \langle |v_r| \rangle^{(i)}, \langle \tau_t \rangle^{(i)}\}$, where $\langle \cdot \rangle ^{(i)}$ denotes an average over all events exhibited by individual $i$.  In a population, phenotypic traits can be described by a distribution $P(\chi)$ that governs the probability that an individual has a specific value for each trait $\chi$.   The distributions $P(\chi)$ for the founding strain and individuals isolated after \num{5}, \num{10} and \num{15} rounds of selection are shown in Fig. \ref{fig_experiment_variance}(A-C).  We quantify phenotypic fluctuations, or cell-to-cell variation, by the standard deviation across the population in each trait, for $N$ cells this is computed as: $\sigma_{\chi} = \sqrt{\frac{1}{N} \sum_i (\chi^{(i)} - \langle \chi^{(i)} \rangle)^2}$.  We note that $\sigma_{\chi}$ describes phenotypic variation driven by both genetic and non-genetic variation in the population except in cases of clonal populations, where $\sigma_\chi$ is due to non-genetic effects alone.

To experimentally quantify phenotypic fluctuations we computed average run durations, tumble durations and run speeds on a per cell basis.  Explicitly, if cell $i$ executes $M$ runs during the \num{5} minutes of tracking we compute $\langle \tau_r \rangle ^{(i)} = \frac{1}{M} \sum_{j=1}^M \tau_{r,j}$.  To quantify the cell-to-cell variation we then compute the standard deviation across individuals $\sigma_{\langle \tau_r \rangle}$.  We compute identical statistics for the tumble duration $\tau_t$ and the run speed $|v_r|$ for founding populations and populations isolated after \num{5}, \num{10} and \num{15} rounds of selection.  Fig. \ref{fig_experiment_variance}(D-F) shows the standard deviations across the population ($\sigma_{\chi}$) for $\chi \in \{\langle \tau_r \rangle, \langle \tau_t \rangle, \langle |v_r| \rangle \}$, indicating a significant decline in the cell-to-cell variation during selection.  In particular, we observe a significant decline between founding population and rounds \num{10} and \num{15} for all phenotypic parameters.  We conclude that selection for faster migration results in reduced phenotypic fluctuations in the population.
% HERE.

The common interpretation for the utility of phenotypic variation is that it may increase survival probability under environmental changes by providing variation with every generation as opposed to genetic mutations which occurs less frequently\cite{Frankel:2014fb,Balaban:2004bq}.  Whether populations are shaped more by phenotypic varation or genotypic variation depends on the degree of phenotypic variation and on the strength and types of environmental selection.  Is this reduction a special feature of the experiment, or can it be understood from general principles?  To address this, we describe below an abstract computational model which is independent of the mechanistic details of our particular experiment.  We ask how the process of iterated selection, whereby cells from the tail of a phenotypic distribution are propagated to the next round, alters cell-to-cell variation.  Our goal with the simulation is to predict how the evolution of cell-to-cell variation depends on the strength of selection.

\section{Abstract model of directed evolution of phenotypic fluctuations}
%%%%%%%%%%%%%%%%%%%%%%%%%%%%%%%%%%%%%%%%%%%%%%%%%%%%%%%%%%%%%%%%%%%%%%%%%%%
% More introductory descriptions for motivation may be added.
%%%%%%%%%%%%%%%%%%%%%%%%%%%%%%%%%%%%%%%%%%%%%%%%%%%%%%%%%%%%%%%%%%%%%%%%%%%
The genotype-phenotype map determines the phenotype of an organism with a given genotype.  How phenotypic selection is coupled to genetic variation is an important question whose answer illuminates fundamental questions such as the evolutionary rate and the evolvability of organisms.  In general, this mapping is a multi-dimensional function that is governed by complex biological features such as gene regulatory and metabolic networks.  As such, in laboratory-based directed evolution experiments the evolutionary dynamics of a specific phenotype are difficult to understand in terms of genetic variation alone. Therefore, we seek a framework that does not rely on an explicitly modeled mapping from genotypes to phenotypes.  For simplicity, we present a computational model of adaptation of a single effective phenotype and its associated genotype, representing a projection of a multi-dimensional phenotype/genotype evolving under selection.  The idea is related to previous population genetics models \cite{lande1976}, but instead of assuming continuous selection due to an assumed fitness landscape, we specify selection through a population bottleneck that is decoupled from the rate of growth.  We use this model to calculate the evolution of phenotypic variation under selection.  The model is necessarily stochastic in order to capture the dynamics of fluctuations.  We do not specify any explicit mechanism for genotype-phenotype mapping or how its functional form changes during evolution.  Instead, phenotypes are random numbers generated from a Gaussian mapping function whose mean is identified with a genotype and whose variance reflects phenotypic fluctuations across individuals with that genotype.  The mean and variance change in evolutionary processes such as point mutations.  Contrary to the conventional population genetics argument for directed evolution that predicts decrease in the variance of phenotype as a result of avoiding deviation from the peak in the fitness landscape, we attempt to understand how the various factors can affect the evolutionary trajectory in a minimal and general model.

\subsection{Main features of the abstract model}
Our abstract model captures key features of a fully realistic model built on a lower-level description such as gene expression.  The main experimentally-relevant factors considered in this abstract model include strength of bottleneck selection, mother-daughter correlation and mutations.  The mother-daughter correlation (or epigenetic inheritance) describes the degree of gene expression level that is passed on to descendants in the absence of mutations and determines how well preserved a phenotype is in subsequent generations.  Mutations stochastically induce changes in the phenotype ($\chi$).  We focus on the effect of the strength of bottleneck selection and the mutation rate.  The correlation between mother and daughter is effective in accumulating phenotype changes during directed evolution if the correlation is high.  However, measurements showed that the mother-daughter correlation is around $0.5$ (data shown in Fig. S1) and is the same for both the founder and the evolved strains.  Moreover, it was shown in the relaxation experiment in \cite{Fraebel:2017wx} that after about 140 generations of growth in well-mixed liquid conditions no additional mutations occurred and the fast migrating phenotype was retained throughout this extended growth period.  Therefore, we set the mother-daughter correlation to not evolve in the directed evolution in our model.   Finally, the bottleneck selection is applied in the trait space instead of the real space, and therefore the details of the experiment including the consumption of nutrients and the process of chemotaxis are not explicitly represented in the model in order to reduce complexity.
%Further, the details of the experiment including the consumption of nutrients and the process of chemotaxis are not explicitly represented in the model but are captured in the process of selection.

%%%%%%%%%%%%%%%%%%%%%%%%%%%%%%%%%%%%%
%\subsection{Phenotype threshold}
In addition, traits such as run speed cannot physically evolve to infinitely large values and thus should be bounded by a threshold $\chi_c$.  The threshold on phenotype represents a limitation of the corresponding cellular machinery, and therefore it fluctuates between cells in general.  
%In our simulations we include upper thresholds on $\chi^{(i)}$ for each individual and on the mean phenotype $\mu_\chi$ of each genotype, and the threshold values for both are set to be random numbers generated for each individual from $\mathcal{N}(\mu_{\chi_c},\eta^2_{\chi_c})$.  
%Due to the threshold, both the distribution of isogenic fluctuations and the distribution of genotype variations are set to be truncated normal distributions.  
Due to the threshold, the trait $\chi$ converges when the mean of phenotype of the population gets closer to the threshold, and therefore the effective evolutionary rate decreases.  We also anticipate that the convergence to the threshold can lead to skewness in the phenotype distribution, because fluctuations in the phenotype cannot exceed the threshold.
%As a result, the range over which the mean trait $\mu_\chi$ and the trait $\chi$ can evolve becomes smaller when $\mu_\chi$ gets closer to the threshold, and therefore it automatically develops an effective \lq\lq slowing down" of the rate of evolution.  In the simulation results shown below the assumption of a threshold leads to skewness in phenotype distributions that has been observed in the experimental data, even though we do not assign any specific functional form for the asymmetry.
%The reason that we do not focus on lower thresholds of phenotype is because it does not matter for directed evolution that evolves in the direction of larger phenotype.  If the directed evolution is designed to evolve in the opposite direction then it is the lower bounds of phenotype that determines the evolution of phenotypic fluctuations.  The model can also be extended to the case where the selected trait depends on two or more phenotypes, and the overall threshold would be determined by the combination of higher or lower thresholds of each trait.  We assumed that the timescale for changes in the threshold $\mu_{\chi_c}$ is very long and set this to be a constant in all simulations.  Therefore the value of the phenotype $\chi$ for a particular genotype $g$ is distributed with a truncated normal distribution with an upper bound which is approximately $\mu_{\chi_c} \pm \eta^2_{\chi_c}$.  We set $\mu_{\chi_c}=100$ and $\eta_{\mu_\chi}=3$.

\subsection{Na\"ive prediction for the effects of variation in selection strength}\label{sec_naive_argument}
We expect that one of the relevant control parameters is the strength of the population bottleneck selection.  We note that in the simulation multiple genotypes can coexist in the population at variable frequencies.  Intuitively, without any physically-determined threshold on phenotype $\chi$, individuals who evolve higher mean trait value and larger phenotypic fluctuations of their genotypes are expected to preferentially populate the right-most tail of the population trait distribution, and so will have a higher probability to be selected.  Therefore, after population amplification where selection is absent, the overall phenotypic variance in the population would be expected to increase monotonically.  However, once the population trait distribution approaches the threshold, the mean trait value of the genotypes of the selected individuals gets close to the threshold, and mutants with similar mean but different phenotypic fluctuations can arise.  If the selection strength is strong, genotypes with both large or small isogenic fluctuations can both contribute large phenotype values near the threshold and be selected, so that the phenotypic variation would be expected to increase.  
On the other hand, genotypes with large isogenic fluctuations will have significant weighting at smaller $\chi$, and therefore if selection strength is not strong enough, genotypes with smaller isogenic fluctuations are more likely to be selected, potentially leading to a decrease in the overall variance.  

These na\"ive and intuitive arguments, however, do not account for the effects of individual variations in threshold, variations of threshold from generation to generation, and the effects of mutations.  Our simulation results, described below, reveal a more subtle and complex series of outcomes in the evolution of phenotypic fluctuations.  As a result, accurately predicting the dynamics requires stochastic quantitative models, and cannot be reliably carried out with na\"ive arguments.

%%%%%%%%%%%%%%%%%%%%%%%%%%%%%%%%%%%%%

\subsection{Detailed description of the abstract model}
In our abstract model, each individual $i$ is represented by a random phenotype value $\chi^{(i)}$ which is determined by the individual's genotype $g$.  $\chi^{(i)}$ is generated from a normal distribution $P(\chi)$ whose mean is $\mu_{\chi}(g)$ and whose variance is $s_{\chi}^2(g)$ in the absence of mother-daughter correlations.  This abstracted phenotype is intended to represent any observable phenotypic variable.  We assume that the phenotype does not change within the individual's lifetime.  In our abstract model of directed evolution, the phenotypic trait $\chi$ is not explicitly stipulated.  Instead, our abstract model is intended to explore the dynamics of phenotypic evolution under generic assumptions about how traits are passed between generations and respond to mutations. 

Individuals reproduce and the offspring acquire mutations with probability $\nu$, causing the daughter's genotype $g'$ to be distinct from the mother's ($g$). Therefore, the daughter's phenotype follows another normal distribution with distinct mean $\mu_{\chi}(g')$ and distinct variance $s_{\chi}^2(g')$. 

In the absence of mutations (i.e. within a clonal population derived from a single genotype $g$), the phenotypes of each new cell are generated based on a bivariate gaussian distribution $P(\chi^{(i)},\chi^{(i')})$ with mother-daughter correlation coefficient $\rho$ that captures the fact that daughter cells have phenotypes  $\chi^{(i')}$  which is correlated with those of their mother $\chi^{(i)}$ \cite{jensen2000}.  Phenotypic correlations between generations in clonal populations can arise from protein copy number fluctuations or non-genetic changes in gene expression \cite{raser2005,feinberg2010}.  For an individual $i',$ which results from fission of individual $i$, its phenotype $\chi^{(i')}$ follows the conditional distribution of the variable $\chi^{(i')}$, given a known value of $\chi^{(i)}$ \cite{jensen2000}:
\begin{equation} \label{bivariate_conditional}
P(\chi^{(i')}|\chi^{(i)})\sim\mathcal{N}(\mu_{\chi}(g)+\rho(\chi^{(i)}-\mu_{\chi}(g)),(1-\rho^2)s^2_{\chi}(g)),
\label{bivariate_conditional}
\end{equation}
where $\mathcal{N}(\mu,s^2)$ is a normal distribution with mean $\mu$ and variance $s^2$.
%
%Under evolution with mutation rate $\nu$, the mean and variance of $\widetilde{P}(\chi)$ change correspondingly, because a mutation affects the distribution of a particular trait.

We calculate these dynamics, along with the procedure for directed evolution through selection, as follows: 

\begin{enumerate}
	\item $N_s$ individuals from a single genotype $g=g_0$ are generated from $P(\chi)=\mathcal{N}(\mu_{\chi}(g_0),s^2_{\chi}(g_0))$, as illustrated in Fig.\ \ref{fig_abstract_illustration}(A).  These $N_s$ clonal individuals are defined as the founder strain, which by construction is a population with a normal distribution of different phenotypes.  
	\item \label{step_reproduce} Each individual with phenotype $\chi^{(i)}$ creates a new individual with phenotype $\chi^{(i')}$.  The new individual mutates to a new genotype $g=g_1\neq g_0$ with a rate $\nu$:  
	\begin{enumerate}
	\item \label{step_mutation} If it mutates, $\chi^{(i')}$ is generated from $P(\chi)=\mathcal{N}(\mu_{\chi}(g_1),s^2_{\chi}(g_1))$, where $\mu_{\chi}(g_1)$ and $s_{\chi}(g_1)$ are generated from $\mathcal{N}(\mu_{\chi}(g_0),\eta^2_{\mu_\chi})$ and $\mathcal{N}(s_{\chi}(g_0),\eta^2_{s_\chi})$ respectively.  The variances $\eta^2_{\mu_\chi}$ and $\eta^2_{s_\chi}$ are assumed to be constant for all parent genotypes ($g_0$). 
	\item \label{step_no_mutation} If the new individual does not mutate, $\chi^{(i')}$ updates based on Eq.(\ref{bivariate_conditional}).
	\end{enumerate}
	An example of the relationship between different phenotypes and the reproduction process is shown in Fig. \ref{fig_abstract_illustration}(B-C).	During reproduction we neglect the degradation of individuals, and thus the population doubles after one generation.  Each individual in the doubled population generates a new individual in the next generation following step \ref{step_mutation} or \ref{step_no_mutation}.  We assume that the mother-daughter correlation ($\rho$) does not evolve.  After $m$ generations, selection is applied to the whole population with $N_f=N_s\times 2^m$ individuals.    
	
	\item \label{step_selection}To apply selection, $N_r$ individuals with the  largest $\chi$ values are chosen from the population.  The selection fraction $N_r/N_f$ is defined to be the selection strength.  $N_s$ individuals are further randomly selected from the $N_r$ individuals to be the seed population for the next round.  {$N_r$ is analogous to the outer edge population sampled with a pipette in the experiments, and $N_s$ represents the individuals that are used to inoculate the new plate.  In experiments, $N_f\sim 10^{10}, N_r\sim 10^8$ and $N_s\sim 10^6$.  Thus, in the bacterial chemotaxis experiments $N_f\gg N_r\gg N_s$.}  
	
	\item In the new round, step (\ref{step_reproduce}) and (\ref{step_selection}) are repeated for the $N_s$ individuals from the previous round.  
	
	\item The phenotypic variance in $N_s$ individuals at the end of each round is measured by growing a population to $N_l=N_s\times 2^l$ individuals by repeating step \ref{step_no_mutation} without mutations.  This mimics the experimental process of single cell tracking in liquid media, where populations are amplified by growth in well-mixed liquid conditions and presumably mutations can be neglected.
\end{enumerate}
	
The parameters in the simulations are: $N_s=100, m=10, l=10, \mu_\chi(g_0)=40, s_\chi(g_0)=8, \eta_{\mu_\chi}=3, \eta_{s_\chi}=1$, with $\nu=0.2$.  Simulations were run over $120$ rounds.  The stochastic values of $\chi$, $\mu_\chi$ and $s_\chi$ are binned to create finite differences between trait values.  The bin sizes in the simulations are $1$, $3$ and $1$ respectively. 
%The parameters were chosen so that $\chi$ values measured in the amplified population of $N_l$ individuals converged in the last $10$ rounds of selection in order to facilitate comparison of observed trends with experiments.  
The selection process is described in Fig. \ref{fig_abstract_illustration}(D).

These simulations do not directly stipulate how the phenotypic fluctuations within a given genotype $s_{\chi}(g)$ evolve -- e.g. these can increase or decrease relative to the parent genotype $g$.  This is intended to avoid any bias on phenotypic fluctuations with respect to the evolving mean trait values.  For example, we do not explicitly stipulate that $s_{\chi}(g)$ decreases as $\mu_{\chi}(g)$ increases.  However, a mechanistic link between the mean and variance of a phenotypic trait could occur in more realistic situations where traits are constrained by trade-offs.  For example, there is usually a fitness cost for a trait to deviate far from the mean, especially when the mean trait values are already optimized for a given environment.  

%In addition, traits such as run speed cannot physically evolve to infinitely large values and thus should be bounded by a threshold $\chi_c$.  The threshold on phenotype represents a limitation of the corresponding cellular machinery, and therefore it fluctuates between cells in general.  
In the abstract model, the effect of threshold is included by considering an upper bound on $\chi^{(i)}$ for each individual and on the mean phenotype $\mu_\chi$ of each genotype, and the threshold values for both are set to be random numbers generated for each individual from $\mathcal{N}(\mu_{\chi_c},\eta^2_{\chi_c})$.  
Due to the threshold, both the distribution of isogenic fluctuations and the distribution of genotype variations are set to be truncated normal distributions.  As a result, the range over which the mean trait $\mu_\chi$ and the trait $\chi$ can evolve becomes smaller when $\mu_\chi$ gets closer to the threshold, and therefore it automatically develops an effective \lq\lq slowing down" of the rate of evolution.  %In the simulation results shown below the assumption of a threshold leads to skewness in phenotype distributions that has been observed in the experimental data, even though we do not assign any specific functional form for the asymmetry.
The reason that we do not focus on lower thresholds of phenotype is because it does not matter for directed evolution that evolves in the direction of larger phenotype.  If the directed evolution is designed to evolve in the opposite direction then it is the lower bound of phenotype that determines the evolution of phenotypic fluctuations.  The model can also be extended in principle to the case where the selected trait depends on two or more phenotypes, and the overall threshold would be determined by the combination of higher or lower thresholds of each trait.  We assumed that the timescale for changes in the threshold $\mu_{\chi_c}$ is very long and set this to be a constant in all simulations.  Therefore the value of the phenotype $\chi$ for a particular genotype $g$ is distributed with a truncated normal distribution with an upper bound which is approximately $\mu_{\chi_c} \pm \eta^2_{\chi_c}$.  We set $\mu_{\chi_c}=100$ and $\eta_{\mu_\chi}=3$.

\subsection{Results of simulations}\label{sec_simulation_result} 
Fig. \ref{fig_abstract_simulation}(A-B) shows the evolution of the distribution of $\chi$ in the amplified population of the $N_l$ individuals after each selection round, denoted by $P_p(\chi)$, under strong selection (panel A) and weak selection (panel B). The phenotypic fluctuation (or the cell-to-cell variation), given by $\sigma_\chi$, is defined as the standard deviation of $P_p(\chi)$, and the average phenotype, represented by $\overline\chi$, is the expectation value of this distribution.
In the remainder of this section, we will walk through the results of the simulations in detail, because there are a number of distinct cases that need to be presented.  In the following section, we will interpret the outcome in terms of the behavior of the isogenic phenotype distributions of genotypes.

The evolution of $\sigma_\chi$ and $\overline\chi$ are shown in Fig. \ref{fig_abstract_simulation}(C-E).  Before round $40$, the simulation results are broadly consistent with our na\"ive prediction, where strong and weak selection leads respectively to increase and decrease in $\sigma_\chi$.  Specifically, Fig. \ref{fig_abstract_simulation}(A) shows an example of strong selection, where the selection strength is the strongest, defined here to be the case that the individuals with the top $N_s$ largest phenotypes are selected ($N_r=N_s\ll N_f$).   In this case $P_p(\chi)$ quickly evolves to large $\chi$ and becomes wider before round $5$, and $\overline\chi$ and $\sigma_\chi$ increase accordingly as shown by the green curves in Fig. \ref{fig_abstract_simulation}(D) and (E) which are averaged over $20$ realizations.  After $P_p(\chi)$ approaches $\mu_{\chi_c}$ around round $5$, $P_p(\chi)$ evolves slower and becomes left-skewed and slightly narrower, but still remains wider than the founder distribution, indicating saturating $\overline\chi$ and a slight decrease in $\sigma_\chi$ which is still larger than the variance in the founder strain as shown in Fig. \ref{fig_abstract_simulation}(D) and (E).  However, the increase in the population variance $\sigma_\chi$ is because of selection of large $s_\chi$ (reflected by increasing $\overline{s}_\chi$ in the green and orange curves in Fig. S3(B)) near the threshold, instead of due to the selection of both large and small $s_\chi$ as would be predicted by the na\"ive argument in Section \ref{sec_naive_argument}.  We note that even though we do not assign any specific functional form for the asymmetry, but only assume truncation to Gaussian isogenic fluctuation distributions, the randomness in threshold automatically leads to smooth and skewed distributions like those we observed in the experimental data.

On the other hand, the case under weak selection, where $N_r/N_f=0.5$ in Fig. \ref{fig_abstract_simulation}(B), shows a different evolutionary trend from the case under strong selection, as would be predicted by the na\"ive argument.  Fig. \ref{fig_abstract_simulation}(B) shows an example where $N_r/N_f=0.5$.  In the simulation, it takes longer (about $15$ rounds) for $P_p(\chi)$ to approach $\mu_{\chi_c}$ and to increase its width.  Similarly, $\overline\chi$ saturates more slowly, and $\sigma_\chi$ only evolves to slightly larger values, as shown by the blue curves in Fig. \ref{fig_abstract_simulation}(D) and (E) respectively.  The increase in $\sigma_\chi$ is due to a different reason than in the case under strong selection as above: here both genotypes with large and small $s_\chi$ can be selected because selection is weak, leading to almost unchanged $\overline{s}_\chi$ (the blue curve in Fig. S3(B)).

In addition, in our numerical simulations we found a scenario which is not predicted by the na\"ive argument.  If the selection strength is neither very strong nor weak, but has a model-dependent intermediate value, the variance initially increases but decreases later due to the accumulation of random mutations.  This can lead to very different evolutionary trajectories from one simulation realization to another (orange curves after round $40$ in Fig. \ref{fig_abstract_simulation}(C)), and thus the population variance $\sigma_\chi$ can either increase or decrease depending in an unpredictable way on the selection round (orange curve in Fig. \ref{fig_abstract_simulation}(E)).  

The final average $\sigma_\chi$ evolving after $120$ rounds is shown in Fig. \ref{fig_abstract_simulation}(F) as a function of selection strength, with different sample population ($N_s$) and generation numbers during population amplification ($m$).  The weaker the selection strength is, the smaller the final $\sigma_\chi$ becomes, because the probability of mutants with small $s_\chi$ being selected is higher.  Similarly, larger sample population and more generations during population amplification allows more mutations with small $s_\chi$ to accumulate in the population and thus leads to smaller $\sigma_\chi$.  Except for the cases under very strong selection (e.g. $N_r \leq 2N_s$ or $N_r/N_s \leq 1/2^9$), the final $\sigma_\chi$ after many selection rounds declines compared with the standard deviation in $\chi$ of the founder strain which is represented by the red dashed line.

We also observed that if the traits are not bounded by a threshold, \textit{i.e.} as \ $\mu_{\chi_c} \rightarrow \infty$, the traits evolve without bound in the simulations.  Accordingly there is no saturation and there is no saturation of trait value after repeated rounds of selection, and there is no decline in the variance in the population.  
We note that besides selection strength the result can also depend on other parameters.  For instance, if the mother-daughter correlation is high or the mutation range of the isogenic fluctuations ($\eta_{s_\chi}$) is small, $s_\chi$ does not mutate enough to increase much while $\mu_\chi$ still evolves to the threshold, and therefore $\sigma_\chi$ can remain small even under very strong selection in this case.

In conclusion, through the simulations of this abstract model for directed evolution we have shown that an upper bound of phenotype can lead to finite-time saturation of the evolving phenotype, and to the decrease of cell-to-cell variation under temperate selection with typical parameter values.  In the case with strong selection, the decrease of cell-to-cell variation is not a necessary consequence of the directed evolution procedure.  Under strong selection, genotypes with large phenotypic fluctuations are favored, and the average phenotype and genotype values increase faster (Fig. \ref{fig_abstract_simulation}(D)).  In this sense, strong selection can be regarded as increasing the evolvability.  In other words, whether phenotypic variation is advantageous or unfavorable depends on the selection strength and constraints on the phenotype.

\subsection{Heuristic interpretation of the simulation results}
%\subsection{Heuristic argument for interplay between selection, mutation and threshold}
Now that we have described the simulation results, we interpret them heuristically in terms of the isogenic phenotype distributions of genotypes.  We emphasize that this is a \textit{post hoc} rationalization of what the simulations revealed, and we cannot simply predict \textit{a priori} these outcomes from na\"ive arguments.
%We expect that one of the relevant control parameters is the selection strength $N_r/N_f$.  We note that in the simulation multiple genotypes can coexist in the population at variable frequencies.  We denote the distribution of phenotypes for the entire population as $P_p(\chi)$.  Now we present some intuitive arguments that try to predict the behavior of $P_p(\chi)$ as a function of selection strength.  
To understand the simulation results, we consider carefully the interplay between selection, mutation and random threshold.  Here we refer to the lower bound on $\chi$ for the selected $N_r$ individuals to be $\chi^*$.  

Without any physically-determined threshold on phenotype $\chi$, the genotypes with larger isogenic fluctuations $s_\chi$ and $\chi_c$ can provide phenotypes with larger $\chi$ at the tail of their distributions $P(\chi)$.  Therefore, under strong selection that acts at the right tail of distributions, the genotypes with larger isogenic fluctuations are more likely to be selected, and the variance of the distribution of phenotypes for the entire population $P_p(\chi)$ after population amplification will increase.  On the other hand, if selection is weak, genotypes with large $s_\chi$ or large $\chi_c$ are not particularly favored, but mutations in $\mu_\chi$ could develop heterogeneity in $P_p(\chi)$ and this leads to an increasing variance.  

When the mean of $P_p(\chi)$ approaches $\mu_{\chi_c}$, $P(\chi)$ becomes truncated by the threshold, as illustrated in Fig. \ref{fig_selection_heuristic}.  Under strong selection, the genotypes with large $s_\chi$ and also large $\chi_c$ contribute the largest $\chi$ in $P_p(\chi)$, and $\mu_{\chi}$ saturates quickly.  When selection is extremely strong, e.g. $N_r\sim N_s$ and $N_r/N_f$ is very small, the selection point $\chi^*$ is very close to the right tail of the dominant genotypes (cyan curve in Fig. \ref{fig_selection_heuristic} (A)).  Before the next bottleneck selection, $\mu_\chi$ of the most mutants is constrained by the random upper bound with an average value $\mu_{\chi_c}$, and therefore mutant genotypes with smaller $s_\chi$ (e.g. purple curve in Fig. \ref{fig_selection_heuristic}(A)) have less density above $\chi^*$ compared with the dominant genotypes.  At the next bottleneck selection, it is extremely unlikely for such a mutation to result in a phenotype in the small interval above $\chi^*$, and thus the final variance remains large since strong selection favors those genotypes with substantial probability density above $\chi^*$.

On the contrary, under weak selection where $N_r/N_f\sim$\num{0.5}, when $P_p(\chi)$ approaches $\mu_{\chi_c}$, the distance between the selection point $\chi^*$ and the average truancation point $\mu_{\chi_c}$ is large (Fig. \ref{fig_selection_heuristic}(B)).  In this case, genotypes with large $s_\chi$ (cyan curve in Fig. \ref{fig_selection_heuristic}(B)) no longer provide high density above the threshold for selection $\chi^*$ and instead have lower probability of exhibiting phenotypes \emph{above} this threshold (shaded cyan area in Fig. \ref{fig_selection_heuristic}(B)).  Therefore selection favors mutants with $\chi^* < \mu_\chi < \mu_{\chi_c}$ and smaller $s_\chi$.  The result is selection for genotypes with smaller $s_\chi$, which leads to higher average $\chi$ and smaller final variance.  If the selection is not very strong or weak, genotypes with larger $s_\chi$ can be more favored at first, but after $P_p(\chi)$ approaches $\mu_{\chi_c}$ the rare mutants with smaller $s_\chi$ can still be selected and have large probability density above the selection point even though they are unlikely to contribute phenotypes at the right tail of $P_p(\chi)$.  Whether the final variance increases or decreases would depend on how many mutants have appeared and fixed.  Since mutations are rare and occur stochastically the final variance of $P_p(\chi)$ is expected to vary as a function of selection strength and the number of selection rounds.

A similar result that phenotypic variation could decrease (or increase) under weak (or strong) directed selection was found in \cite{hill2004} with a restricted bi-allele multi-loci model, for eight rounds, and the overall phenotype fluctuation of the population was assumed to be described by the mean and variance even after selection, and therefore could not capture the skewness effect.  Also the effects of threshold and saturation of traits were not included \cite{hill2004,elowitz2010}. 

In general, a reduction in phenotypic fluctuations could be interpreted as stabilizing selection due to canalization \cite{waddington1942}, but the mechanism in this case is different from ours because there is no explicit threshold present.  In the case of canalization, specific biological buffering mechanisms such as capacitance \cite{lindquist1998} are more likely to be at work.  In short, our simulations suggest an alternative mechanism for phenotypic variation, arising as a generic consequence of bounded phenotypic variation under strong or weak selection.

\subsection{Comparison between the experiment and the abstract model} 

The experimental results show that the variance of the run speed decreases with the
number of rounds of selection, a result that our model predicts to occur when selection is weak.  How can we estimate whether or not our experiment is truly in
the weak selection regime?  A na\"ive measure of the selection strength is the ratio $N_r/N_f$
which we estimate to be order $10^2$ in the experiment.  Does this indicate strong selection then?  It is difficult
to draw a clear conclusion about this because, in general, selection acts on the phenotype space. The selection strength should be defined including the weighting of phenotype values, and not simply the
number fraction that assumes equal weighting of each phenotype.  In our experiment selection was applied in real space on agar plates, and thus the real physical phenotype that is being selected is a compound trait of multiple variables.  Therefore, the selection fraction in the abstract model might not be simply related to the selection
strength in the physical system.  Thus, in order to test how the trend of phenotypic variance evolves with selection strength, it would be necessary to perform another set of experiments with different selection strengths, either a smaller selection fraction or selecting at different part of the population profile, to compare with the current experimental result shown in Fig. \ref{fig_experiment_variance}.  
In addition, the abstract model considers selection and evolution of a low-level trait of individuals instead of an emergent trait at population level.  To explicitly compare with the experiment, we could extend our model by including two or more phenotypes and study the combined effect.  For example, since the selection on colony is applied on the spatial position in our experiment, we may regard the selected property as dominance of length scale, which could be a function of run speed, tumble frequency and growth rate in the case of colony expansion.  
In the experiment, selection is applied on migration of the whole population, which is the property resulted from combined selection of individual chemotaxis and growth between two bottleneck selections.  To explicitly include these features, it will require more variables and parameters, such as nutrient concentration and trait-dependent uptake rate which mimics the selection due to chemotaxis.  These are planned for a future publication. 

%%%%%%%%%%%%%%%%%%%%%%%%%%%%%%%%%%%%%%%%%%%%%%%%%%%%%%%%%%%%%%%%%%%%%%%%%%

\section{Biological mechanisms}
\begin{figure}[!ht]
\begin{centering}
\includegraphics[scale=0.45]{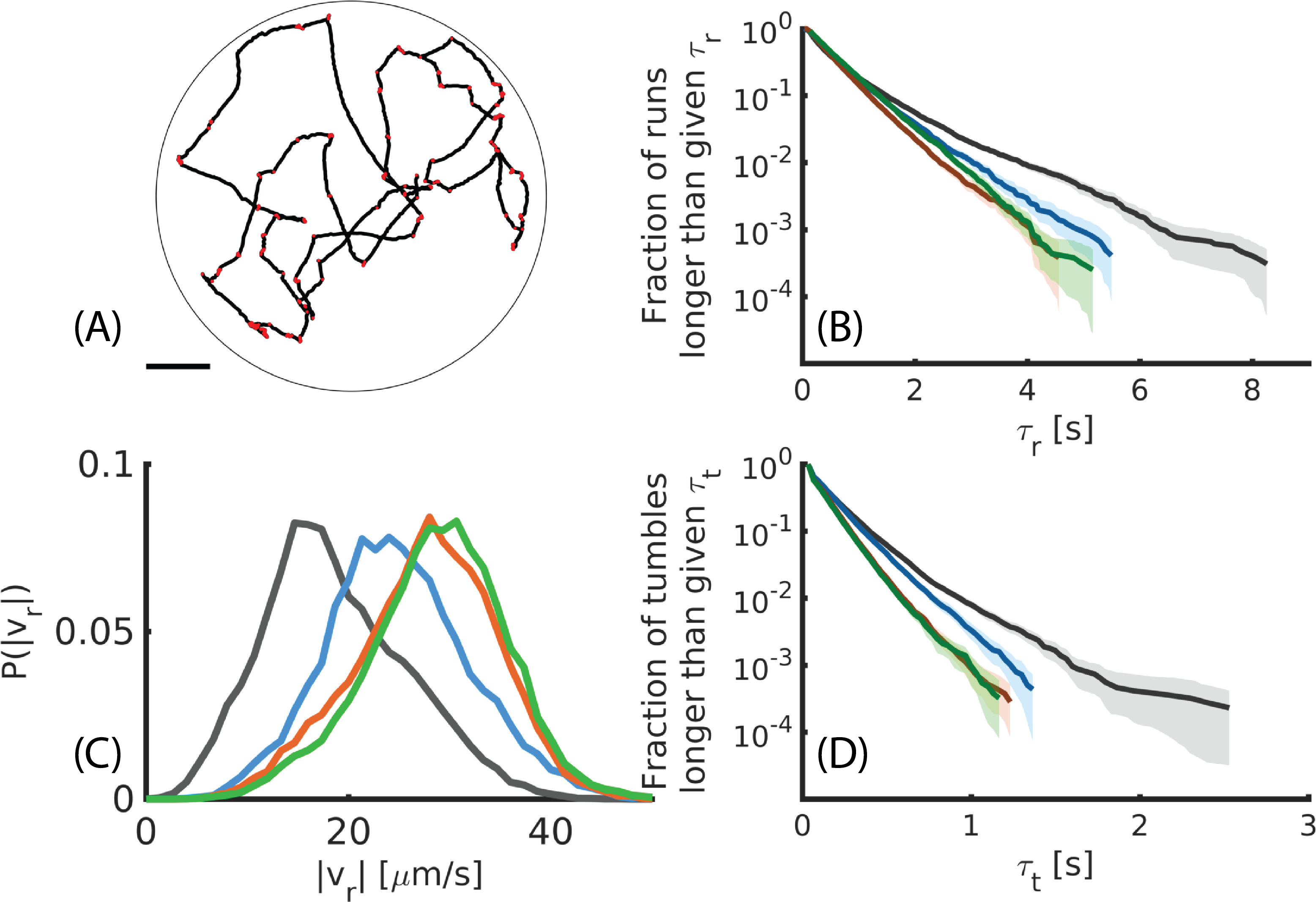}
\caption{\textbf{Dynamics of phenotypic evolution:}  (A) An example \SI{50}{\second} long swimming trajectory for a single cell trapped in a microfluidic chamber.  The boundary of the chamber is shown by the light black circle.  Running events are shown in black and tumble events in red.  Scale bar is \SI{50}{\micro\meter}.  (B) Aggregate complementary cumulative distribution functions of run durations observed from cells isolated prior to selection (founder,black) and after \num{5} (blue), \num{10} (orange) and \num{15} (green) rounds of selection.  Strains tracked were isolated from replicate \num{1} in Fig. \ref{Figure1}.  Distributions are constructed from all run events that were not interrupted by collisions with the chamber boundary for \num{140} (founder), \num{79} (round 5), \num{97} (round 10) and \num{96} (round 15) individuals executing a total of \num{19597}, \num{12217}, \num{18505} and \num{15928} run events respectively.  The mean and standard deviation of run durations are (mean:sd) \SI{0.66}{\second}:\SI{0.78}{\second}, \SI{0.63}{\second}:\SI{0.61}{\second}, \SI{0.58}{\second}:\SI{0.51}{\second} and \SI{0.64}{\second}:\SI{0.57}{\second} respectively.  Shaded regions indicate \num{95}\% confidence intervals from bootstrapping. (C) Distributions of run speeds ($|v_r|$) for the four strains shown in (B), colors from (B) apply.  Distributions are constructed by computing an average speed for each run event.  Means of these distributions are \SI{18.7}{\micro\meter\per\second} (founder), \SI{24.9}{\micro\meter\per\second} (round 5), \SI{27.6}{\micro\meter\per\second} (round 10), and \SI{28.7}{\micro\meter\per\second} (round 15).  The increase in $|v_r|$ is statistically significant between each successive population ($p<0.001$, rank sum test).  (D) Shows the tumble duration distributions for the same four strains shown in panels (B-C).  The mean and standard deviation of tumble durations are (mean:sd) \SI{0.18}{\second}:\SI{0.20}{\second}, \SI{0.17}{\second}:\SI{0.16}{\second}, \SI{0.14}{\second}:\SI{0.13}{\second} and \SI{0.14}{\second}:\SI{0.12}{\second} for founder, round \num{5}, \num{10} and \num{15} respectively.    Data are recapitulated from \cite{Fraebel:2017wx}.\label{fig_experiment_distribution}}
\end{centering}
\end{figure}

\begin{figure}[!ht]
\begin{centering}
\includegraphics[width=0.75\columnwidth]{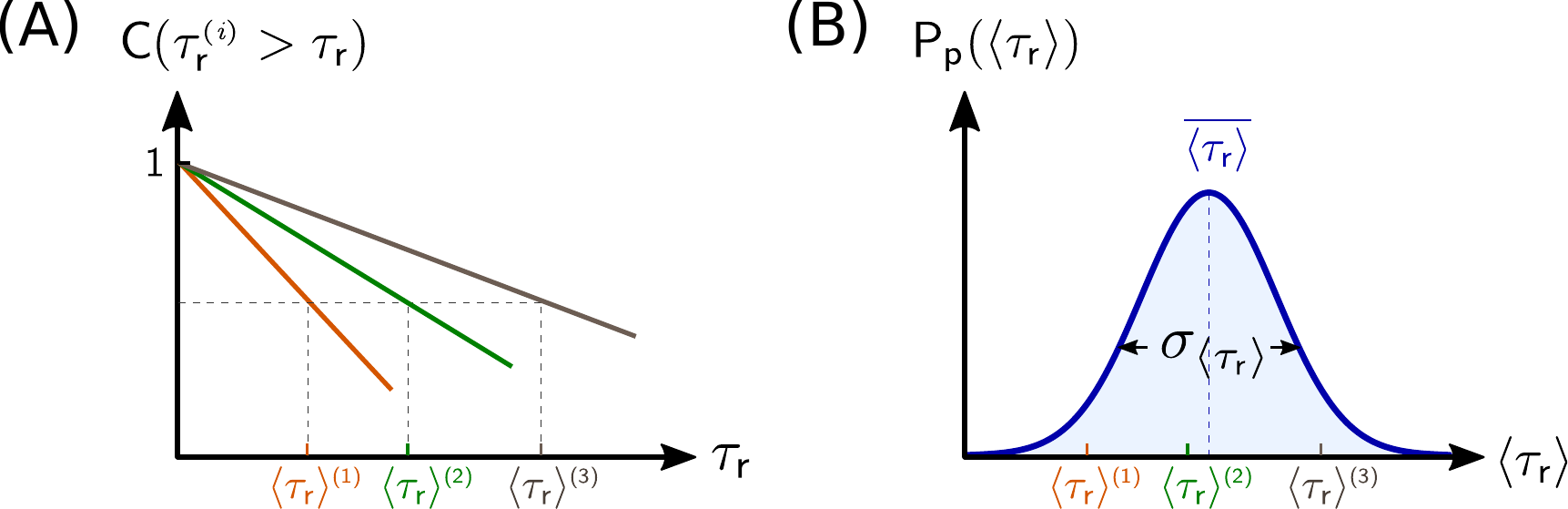}
\caption{\textbf{Illustration of phenotype distributions:}  (A) An example of the complementary cumulative distribution function for run duration from statistics of all run events of different individuals.  The average run duration of individual $i$ ($\langle\tau\rangle^{(i)}$) is read off by fitting the exponential distribution.  (B)  Distribution of individual run duration $\langle\tau\rangle^{(i)}$ in the population generated from (A).  The cell-to-cell variation is characterized by the standard deviation of $P_p(\langle\tau_r\rangle)$, $\sigma_{\langle\tau_r\rangle}$.   \label{fig_illustration_distribution}}
\end{centering}
\end{figure}

Our abstract simulation makes a clear prediction about how phenotypic fluctuations should evolve in the presence of constraints on phenotypes under selection.  Fig. \ref{fig_experiment_distribution} shows that over the course of selection the swimming speed of the cell saturates at approximately \SI{28}{\micro\meter\per\second} and does not change between rounds \num{10} and \num{15} of selection.  This suggests the possibility that $|v_r|$ is in fact bounded from above in a manner similar to our evolutionary simulations.  We note that the precise mechanism of this constraint is not known, but may be hydrodynamic, metabolic or genetic in origin.  For example, the swimming speed increases with flagellar bundle rotation rate\cite{Darnton:2007cta} which depends on the proton motive force and the pH, both of which depend on the metabolic state of the cell.  Swimming speed is also under genetic regulation through a braking mechanism acting on the flagellar motors \cite{Boehm:2010du}.  These mechanisms likely impart an upper bound on the swimming speed of the cell; indeed such a bound must exist given the finite propulsive force supplied by the flagella.  Since we observe a saturation in swimming speed between rounds \num{10} and \num{15} of selection (Fig. \ref{fig_experiment_distribution}(C)) and a concurrent decline in phenotypic fluctuations for $|v_r|$ (Fig. \ref{fig_experiment_variance}) we speculate that this reduction has as its basis a dynamic similar to our abstract model (Fig. \ref{fig_abstract_simulation}), whereby the swimming speed is evolving towards an upper bound.  

While swimming speed ($|v_r|$) appears to evolve towards an upper bound we observe a decline in run durations during selection as well as a decline in the phenotypic fluctuations in $\tau_r$ and $\tau_t$ (Figs. \ref{fig_experiment_distribution} and \ref{fig_experiment_variance}).  It is less clear that explicit bounds apply to run and tumble durations.  Indeed, mutants which exhibit very long or very short run durations have been isolated.  Moreover, phenotypic fluctuations in the temporal statistics of runs and tumbles have been studied in \textit{E. coli} for decades, and the molecular origins of these fluctutions are well understood.  Since the seminal work of Koshland and Spudich \cite{Spudich:1976wv}, we now know that copy number fluctuations of the enzyme \textit{cheR} and \textit{cheB} drive large fluctuations in the run-tumble statistics at the single motor and single cell level\cite{Korobkova:2004vs,Frankel:2014fb,Dufour:2016bo}. Dufour \emph{et al.} \cite{Dufour:2016bo} measured both gene expression and run-tumble statistics in single-cells to show a reduction in phenotypic fluctuations with increasing [\textit{CheR}] and [\textit{CheB}] concentrations \textit{in vivo}.  Phenotypic fluctuations declined when concentrations of both proteins increased while the ratio [\textit{CheR}]/[\textit{CheB}] remained constant \cite{Dufour:2016bo}.  Furthermore, increasing expression of both genes resulted in an increase in tumble frequency precisely as we observe in our selection experiment \cite{Dufour:2016bo}.  In a separate study, Vladimirov \emph{et al.} \cite{Vladimirov:2008ik} show that the expression levels of both \textit{CheR} and \textit{CheB} are higher at the periphery of a colony expanding through \SI{0.27}{\percent} agar than at the center.  Taken together these studies suggest that increasing \textit{CheR} and \textit{CheB} expression should reduce phenotypic fluctuations in $\tau_r$ and $\tau_t$ and that this reduction is correlated with distance from the center of the colony.

	In light of this understanding we examined the mutations present in strains after selection.  We performed whole genome sequencing on the ancestral strain as well as populations isolated after \num{5}, \num{10} and \num{15} rounds of selection for four replicate selection experiments\cite{Fraebel:2017wx}. In every replicate we observe an identical mutation at $>$\num{70}\% abundance by round \num{5} and fixed by round \num{10}: a single nucleotide polymorphism which inserted a stop codon at position \num{185} in the \num{424} residue \textit{ClpX} protein (\textit{clpX}E185*). \textit{ClpX} is the specificity subunit of the \textit{ClpX-ClpP} serine protease, which degrades many target proteins including \textit{FlhDC}.   \textit{flhDC} is the master regulator of a coherent feedforward motif which governs the expression of motility and chemotaxis genes including \textit{cheR} and \textit{cheB} enzymes, which are determinants of phenotypic fluctuations\cite{Kalir:2001gr}. 
	
		To investigate the role of the mutation we observed in \textit{clpX} in phenotypic fluctuations, we reconstructed the \textit{clpX}E185$^*$ mutation in the ancestral background using scarless recombineering.  We confirmed that this mutation alone is sufficient to drive faster migration through increasing run speed and decreasing growth rate \cite{Fraebel:2017wx}.   Moreover, this mutation alone causes a decrease in the phenotypic fluctuations in run duration and tumble duration, but not run speed relative to the ancestral population (Fig. \ref{fig_clpX}).  
	
	We considered whether the mutation we observe in \textit{clpX} might logically result in increased levels of \textit{cheR} and \textit{cheB} and therefore the reduced phenotypic fluctuations we observe.  Previous studies have shown that mutations in \textit{ClpX} increase levels of \textit{FlhDC} in the cell\cite{Tomoyasu:2003dx}.   Zhao \emph{et al.}\cite{Zhao:2007dz} show that deleting \textit{flhDC} results in substantial reduction in expression of the downstream \textit{cheR/B} genes.  However, inducing \textit{FlhDC} expression above wild-type levels appears not to increase expression of downstream genes substantially\cite{Zhao:2007dz}.  Despite this, single cell measurements show a positive correlation between \textit{flhC} and \textit{cheY} expression levels\cite{Lovdok:2008jj}.  Since \textit{cheY} is co-transcribed with \textit{cheR} and \textit{cheB} we speculate that increases in \textit{FlhDC} levels in the cell may drive increases in \textit{cheR} and \textit{cheB} expression and that could reduce phenotypic fluctuations.  Further studies are needed to directly measure the \emph{meche} operon expression levels in the presence and absence of the \textit{clpX} mutation we observe.  Since we cannot stipulate whether expression of the relevant genes is subject to a bound, we cannot conclude that the mechanism proposed in our abstract model describes the decline in fluctuations in run and tumble durations.
	
	The \textit{clpX}E185* mutation alone drives an increase in run speed to \SI{24.2}{\micro\meter\per\second} from \SI{18.2}{\micro\meter\per\second} for the founder whereas the average run speed of the round \num{15} evolved strain is \SI{28.7}{\micro\meter\per\second}\cite{Fraebel:2017wx}.  These results suggest that the mutant run speed is, on average, far from the apparent upper bound in swimming speed.  As our abstract model would predict for the mutant, we observe no decrease in $\sigma_{\langle |v_r| \rangle}$ in the mutant relative to the founder -- potentially because the mutant phenotype is not constrained by an upper bound on run speed.

%Furthermore, measurements of \textit{CheR} and \textit{CheB} expression expression as a function of spatial location in populations expanding through porous agar gels show increased expression levels at the front[CITE-Sourjik].  These results suggest that process of migration through a porous environment selects for cells with reduced phenotypic fluctuations.  Cells with reduced levels of phenotypic fluctuations will exhibit fewer very long runs and therefore have a lower probability of being trapped in the gel [CITE - Berg,Sourjik].  
	
%	One common evolutionary strategy for modulating gene expression is the disruption of negative regulatory elements.  Negative regulatory elements are, for example, transcriptional repressors - proteins that bind DNA and inhibit the expression of downstream genes (regulation of the \textit{lac} operon being the canonical example.)  Several studies have shown that mutations that alter negative regulation, either by disrupting the binding site of the transcriptional repressor, or by loss-of-function mutations in the repressor itself [CITE-Rainey,Tavazoie].  The widespread observation of mutations altering negative regulatory elements is believed to arise from their larger target size relative to gain-of-function mutations.  Recent studies of the evolvability of specific enzymes which exhaustively sample the genotypic space of a single enzyme confirm the overwhelming probability that mutations reduce enzymatic function relative to wild-type [CITE-Rama]

\section{Discussion}
In our measurements we report that selection drives reduction in phenotypic fluctuations associated with chemotactic mobility.  We also identified the mutations that appear to be implicated in this evolution of phenotype fluctuations.  Are the results surprising, or could they have been predicted on general grounds related to the fluctuation-dissipation theorem and other global properties of stochastic gene expression?  

\begin{figure*}[ht!]
\begin{centering}
\includegraphics[width=0.93\columnwidth]{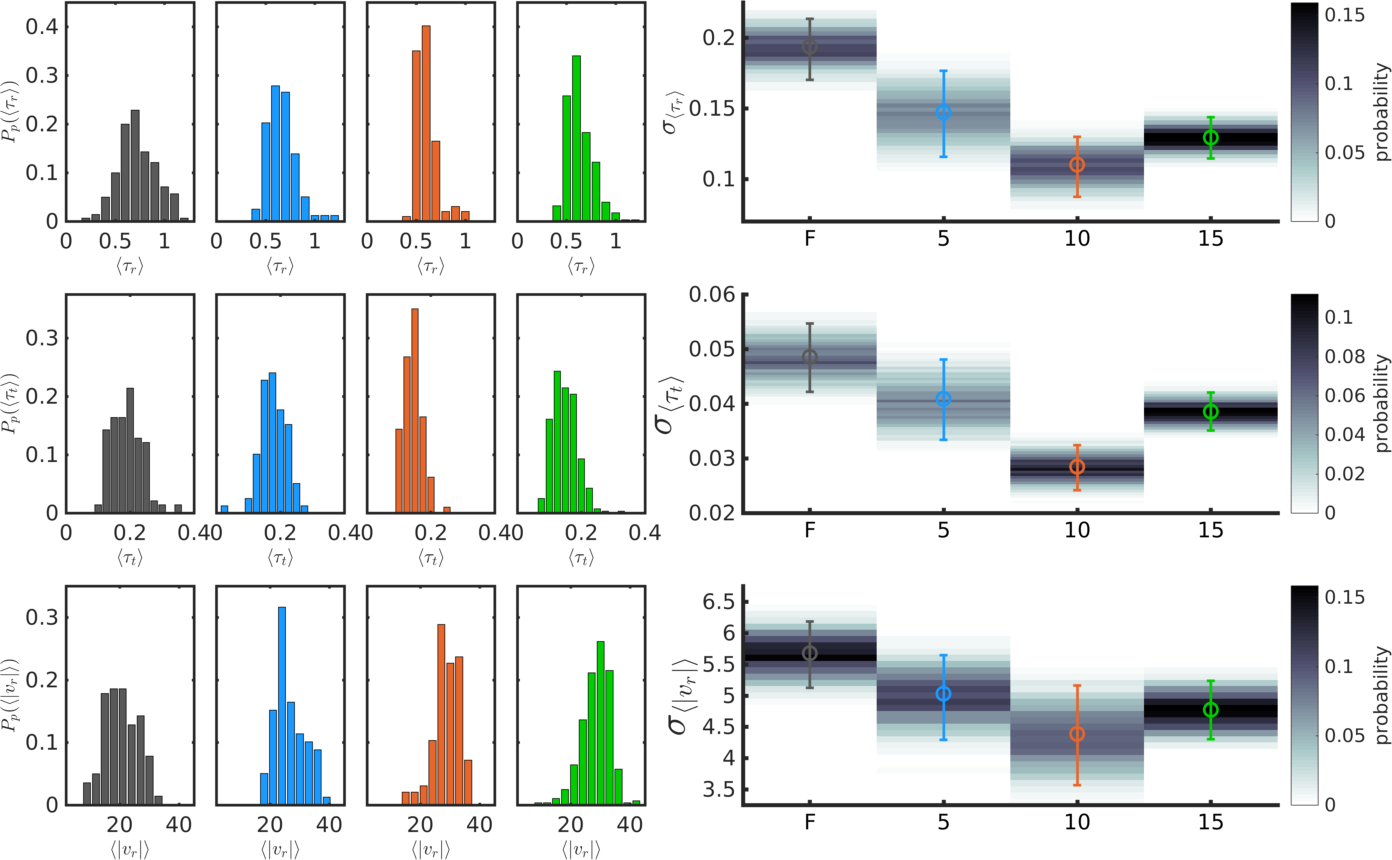}
\caption{\textbf{Cell-to-cell behavioral variation declines with selection:} Phenotype distributions ($P(\chi)$) for (A) $\tau_r$ (B) $\tau_t$ and (C) $|v_r|$ from raw data.  Individuality for evolved populations for (D) $\tau_r$ (E) $\tau_t$ and (F) $|v_r|$.  We compute $\langle \tau_r \rangle$, $\langle \tau_t \rangle$ and $\langle |v_r| \rangle$ for each individual tracked and a standard deviation across individuals for each parameter ($\sigma_{\chi}$).  $\sigma_{\chi}$ was computed for \num{140} cells (founder), \num{79} cells (round 5), \num{97} cells (round 10) and \num{96} cells (round 15).  The circles show the sample $\sigma_{\chi}$ for each population.  \SI{95}{\percent} confidence intervals from bootstrapping for each population are given by the error bars.  Colormap shows the probability distribution of $\sigma_{\chi}$ from bootstrapping.  Note distinct colorbars for each panel.  All populations exhibit a decline in $\sigma_{\chi}$ relative to founder that is significant ($p<0.05$, permutation test) except for $\sigma_{\langle \tau_t \rangle}$ and $\sigma_{\langle |v_r| \rangle}$ in round 5.  The increase in $\sigma_{\chi}$ between rounds \num{10} and \num{15} is only statistically significant for tumble duration ($p =$ \num{5e-4})\label{fig_experiment_variance}}
\end{centering}
\end{figure*}

\begin{figure}[!ht]
\begin{centering}
\includegraphics[width=0.8\columnwidth]%[height=1.6in]
{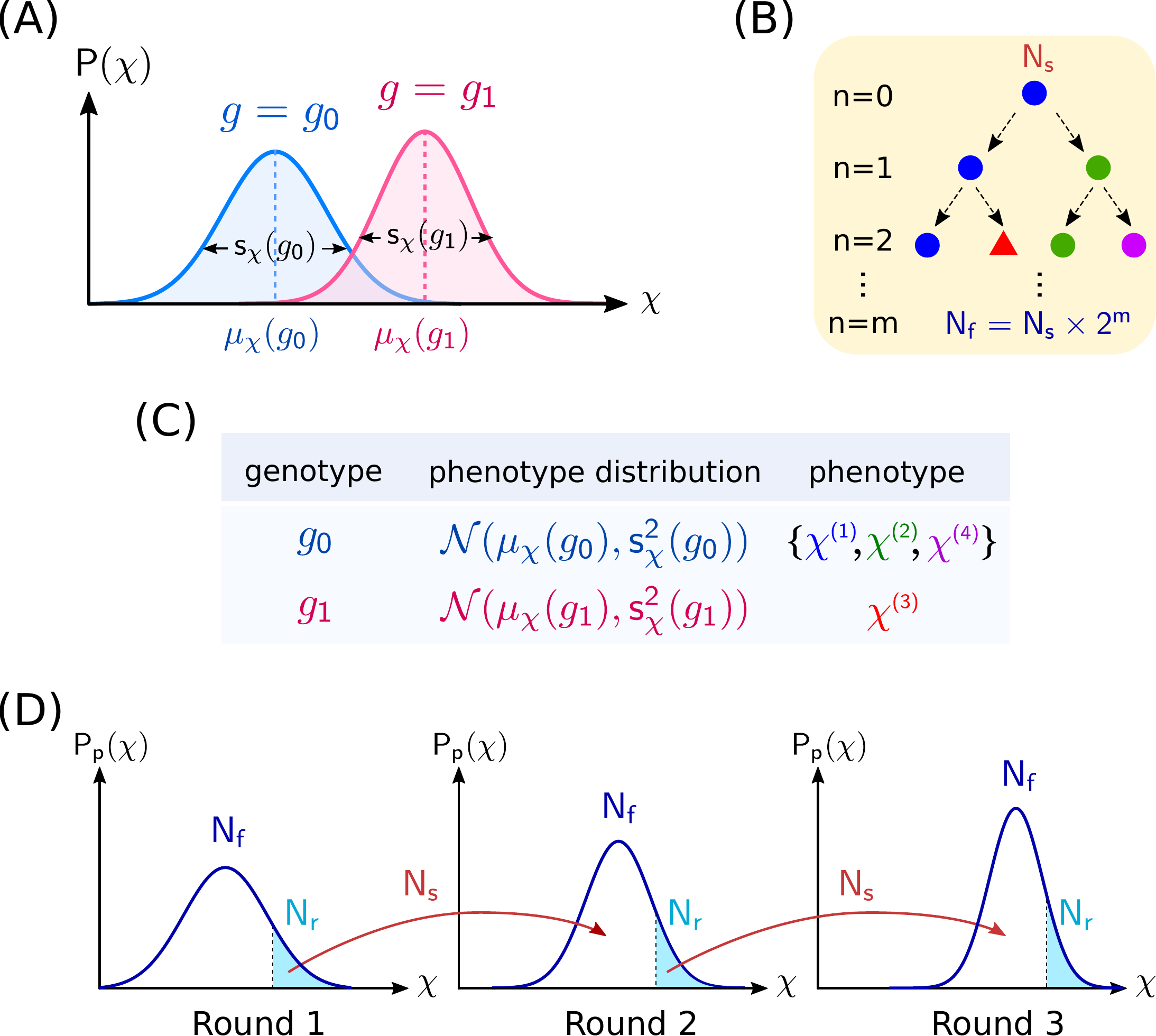}\newline
\caption{\textbf{Scheme of the abstract model.} Illustration of selection procedure (see text for definition of notation): (A)  Phenotype distributions for two genotypes ($g_0,g_1$).  The phenotype of each genotype $g_i$ is described by a normal distribution with mean $\mu_{\chi}(g_i)$ and standard deviation $s_{\chi}(g_i)$.  (B)  Initially the founder strain with $N_s$ individuals whose phenotypes $\chi^{(i)}$ are drawn from $\mathcal{N}(\mu_\chi(g=g_0),s^2_\chi(g=g_0))$ is generated.  The $N_s$ individuals reproduce new individuals in the first round with a mutation rate $\nu$.  For example, for one of the initial $N_s$ individuals with the founder genotype $g_0$ (circle) and a certain phenotype ($\chi^{(1)}$, in blue color) which is determined by mother-daughter correlation based on Eq. (\ref{bivariate_conditional}), its daughter may have the same genotype but different phenotype ($\chi^{(2)}$, in green color) if it does not mutate.  If the daughter mutates, the daughter is assigned a new genotype (triangle) with $\mu_\chi(g_1)$ from $\mathcal{N}(\mu_\chi(g_0),\eta^2_{\mu_\chi})$ and $s_\chi(g_1)$ from $\mathcal{N}(s_\chi(g_0),\eta^2_{s_\chi})$, and its phenotype $\chi^{(3)}$ is drawn from $\mathcal{N}(\mu_\chi(g_1),s^2_\chi(g_1))$ (in red color).  All genotype and phenotype values are truncated by a random upper bound $\chi_c$ chosen from $\mathcal{N}(\mu_{\chi_c},\eta^2_{\chi_c})$.  (C) shows a table of phenotypes ($\chi^i$) and their corresponding genotypes and phenotype distributions.  Note that individuals with the same genotype stochastically differ in their phenotypes (first row).  After $m$ generations of the process shown in (B), the population becomes $N_f=N_s\times 2^m$.  (D) The top $N_r$ individuals are selected from $N_f$ individuals, and $N_s$ individuals are randomly sampled from $N_r$ individuals to start the second round.  In the next round, $N_s$ individuals repeat reproduction steps in (A) until the population reaches $N_f$ again.  How close the average phenotype of $N_f$ in the next round is to the average phenotype of $N_r$ of the previous round depends on how small $m$ is and how high the correlation between mother and daughter is.  At the end of each round, the selected $N_s$ individuals reproduce for $l$ generations without mutations.  These $N_l=N_s\times 2^l$ individuals represent the population of each strain grown in liquid media prior to single-cell tracking.  
\label{fig_abstract_illustration}}
\end{centering}
\end{figure}

\begin{figure}[!ht]
\begin{centering}
\includegraphics[width=0.75\columnwidth]%[height=1.6in]
{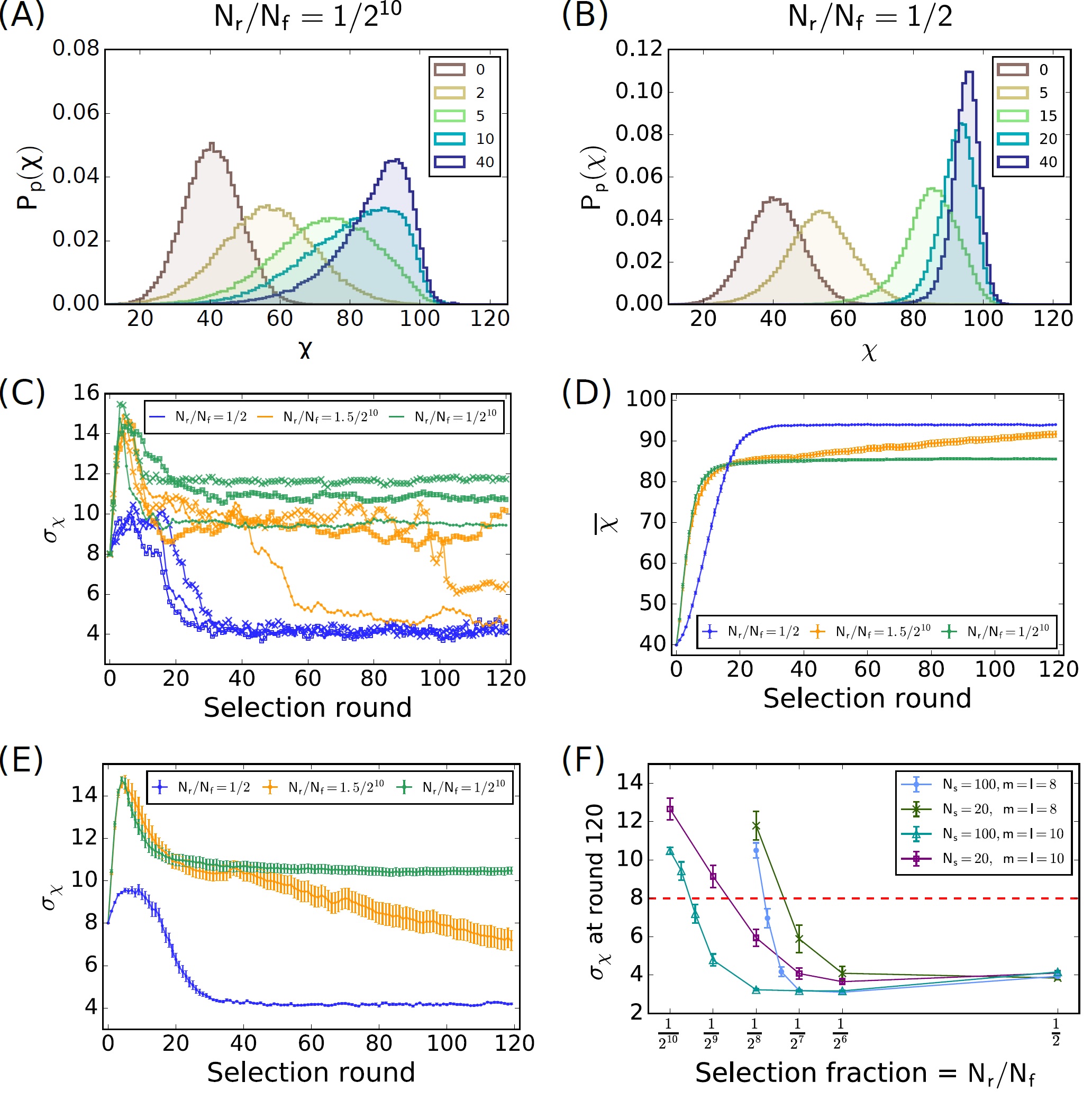}\newline
%
%\caption{\textbf{Simulations of the abstract model} (cation in next page).}
%%\caption[]
%%\end{center}
%%}
%\end{figure}
%\clearpage
%\begin{figure}
%%    \captionsetup{labelformat=adja-page}
%    \ContinuedFloat
\caption{\textbf{Simulations of the abstract model} : 
(A) The distribution of $\chi$ of $N_l$ individuals at different rounds under the strongest selection (e.g.\ the top $N_s$ individuals are selected and $N_r/N_f=1/2^{10}$): the distribution $P_p(\chi)$ quickly evolves and reaches $\mu_{\chi_c}$, and its width remains larger than the founder population even after $P_p(\chi)$ reaches the threshold around round $5$, which implies that the genotypes with smaller $s_\chi$ are not particularly selected under strong selection.  $\mu_{\chi_c}$ and $\eta_{\chi_c}$ are denoted by the vertical and horizontal red line.  
(B) $P_p(\chi)$ under very weak selection (e.g.\ $N_r/N_f=1/2$): when $P_p(\chi)$ is away from $\mu_{\chi_c}$, its width increases when the mean of $P_p(\chi)$ evolves.  The tail of $P_p(\chi)$ reaches $\mu_{\chi_c}$ slower than the case of strong selection (around round $15$), and after that $P_p(\chi)$ becomes tilted and narrower, indicating the overall variance in $\chi$ first increases and then declines eventually.
%(A) The distribution of $\chi$ of $N_l$ individuals at different rounds under weak selection (e.g.\ $N_r/N_f=1/2$): when the distribution $P_p(\chi)$ is away from $\mu_{\chi_c}$, its width increases when the mean of $P_p(\chi)$ evolves, indicating the overall variance in $\chi$ increases.  When the tail of distribution $P_p(\chi)$ starts to reach $\mu_{\chi_c}$ after round $10$, $P_p(\chi)$ becomes tilted and narrower.  $\mu_{\chi_c}$ and $\eta_{\chi_c}$ are denoted by the vertical and horizontal red line.  (B) $P_p(\chi)$ under very strong selection (e.g.\  $Nr/N_f=1/2^{10}$): $P_p(\chi)$ evolves and reaches $\mu_{\chi_c}$ faster, but its width remains larger than the founder population even after $P_p(\chi)$ reaches the threshold, which implies that the genotypes with smaller $s_\chi$ are not particularly selected under strong selection.  
(C) The evolutionary trajectories in three simulation replicate (differentiated by different marker shapes) under different selection strength (blue: $N_r/N_f=1/2$; orange: $N_r/N_f=1.5/2^{10}$; green $N_r/N_f=1/2^{10}$) show very stochastic dynamics due to random mutations.  The overall mean and the standard deviation of $\chi$ are shown in (D) and (E) respectively, averaged over $20$ replicates.  If the selection strength is not extremely strong or weak, the increasing variance can decrease randomly due to accumulated mutations, which can lead to either increase or decrease in the final variance depending on the number of selection rounds (orange curve in (E)).  (F) The overall variance in $\chi$ at round 120 is measured as a function of selection strength.   The final variance is smaller for larger sample population ($N_s$) and more generations between selection rounds ($m$) due to the accumulation of more mutations.   The red dashed line represents the standard deviation of $\chi$ for the founder strain.  Parameters in the above simulations: $\mu_\chi(0)=40, s_\chi(0)=8, \mu_{\chi_c}=100, \eta_{\mu_\chi}=3, \eta_{s_\chi}=1, \eta_{\chi_c}=3$, and the stochastic values of $\chi$, $\mu_\chi$ and $s_\chi$ are binned with bin sizes equal to $1,3$ and $1$ respectively.  In (A)-(E) $N_s=100$ and $m=l=10$.\label{fig_abstract_simulation}}
\end{centering}
\end{figure}

\begin{figure}[!ht]
\begin{centering}
\includegraphics[width=0.8\columnwidth]%[height=1.6in]
{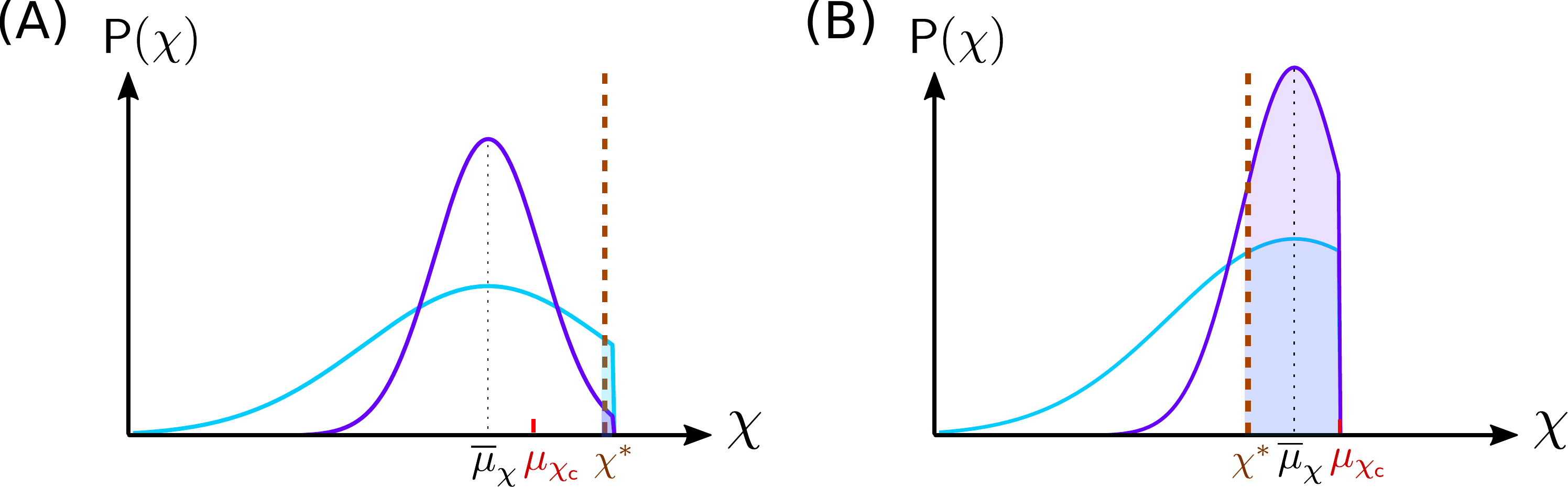}\newline
\caption{\textbf{Heuristic prediction of the cell-to-cell variation in the abstract model.} Illustration of selection when the phenotype distribution $P_p(\chi)$ approaches $\mu_{\chi_c}$ under different strength of population bottleneck selection.  (A)  Under very strong selection where $N_r\sim N_s$, the genotypes with larger isogenic fluctuations $s_\chi$ and $\chi_c$ contribute larger phenotype $\chi$ at the tail of their distributions $P(\chi)$.   Once $P_p(\chi)$ approaches $\mu_{\chi_c}$, $P(\chi)$ becomes truncated by the threshold, and the selction point $\chi^*$ is very close to the right end of the dominant genotypes (cyan curve).  Before the next bottleneck selection, $\mu_\chi$ of the most mutants is bounded by the random threshold with an average value of $\mu_{\chi_c}$, and the mutants with smaller $s_\chi$ (purple curve) have less density above the selction point $\chi^*$ compared with the dominant genotypes.  At the next bottleneck selection, it is extremely unlikely for such a mutation to result in a phenotype in the small interval above $\chi^*$, and thus the final variance remains large since strong selection favors those genotypes with substantial probability density above $\chi^*$.  
(B) On the other hand, if selection is weak, where $N_r/N_f$ is large ($\sim$\num{0.5}), genotypes with large $s_\chi$ or large $\chi_c$ are not particularly favored.  However, mutations in $\mu_\chi$ can develop heterogeneity in $P_p(\chi)$ and thus leads to an increasing variance.   When $P_p(\chi)$ evolves near $\mu_{\chi_c}$, the distance between the selection point $\chi^*$ and the average truancation point $\mu_{\chi_c}$ is large.  Genotypes with large $s_\chi$ (cyan curve) no longer provide high density above the threshold for selection $\chi^*$, but instead have less substantial probability of exhibiting phenotypes \emph{below} this threshold (shaded cayn area).  Further mutants with $\chi^* < \mu_\chi < \mu_{\chi_c}$ and smaller $s_\chi$ are favored under selection.  The result is selection for genotypes with smaller $s_\chi$, which leads to higher average $\chi$ and smaller final variance.  When selection is not very strong or weak, genotypes with larger $s_\chi$ can be more favored at first, but after $P_p(\chi)$ approaches $\mu_{\chi_c}$ the rare mutants with smaller $s_\chi$ can still be selected and have large probability density above the selection point even though they are unlikely to provide phenotypes at the right tail of $P_p(\chi)$.  Therefore in this case, depending on how many mutants have accumulated, the final variance of $P_p(\chi)$ can either increases or decreases and should be a function of selection strength and the number of selection rounds.\label{fig_selection_heuristic}}
\end{centering}
\end{figure}

\begin{figure*}[ht!]
\begin{centering}
\includegraphics[scale=0.45]{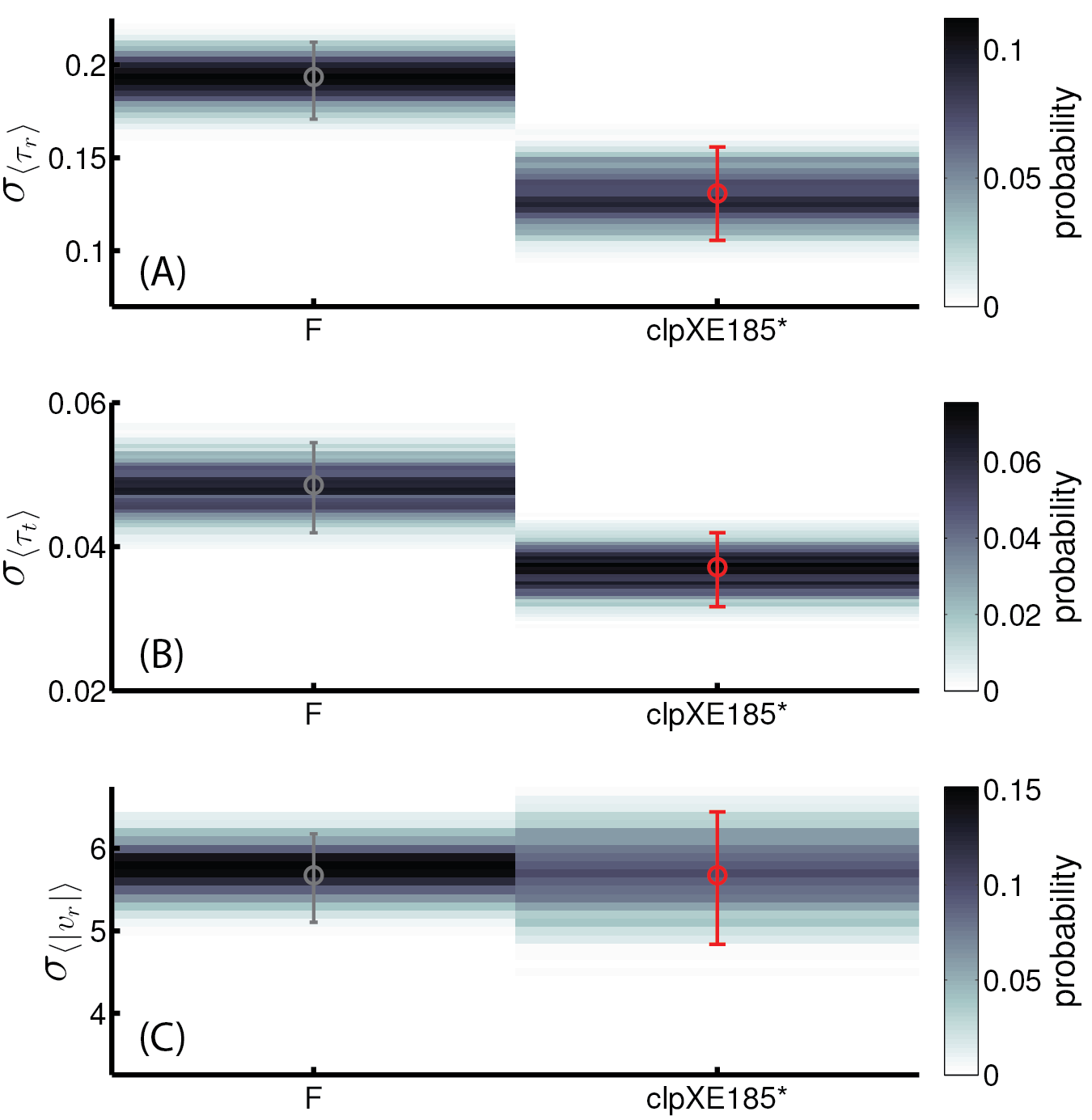}
\caption{\textbf{Cell-to-cell behavioral fluctuations in \textit{clpX} mutant:} Individuality for a mutant with the \textit{clpX}E185$^*$ mutation compared to the founder.  Individuality for each population for (A) $\tau_r$ (B) $\tau_t$ and (C) $|v_r|$.  We compute $\langle \tau_r \rangle$, $\langle \tau_t \rangle$ and $\langle |v_r| \rangle$ for each individual tracked and a standard deviation across individuals for each parameter ($\sigma_{\chi}$).  Data from \num{140} founder cells is reproduced from Fig. \ref{fig_experiment_variance} and compared to \num{82} \textit{clpX}E185$^*$ cells.  Panels are identical to Fig. \ref{fig_experiment_variance} with circles showing the sample $\sigma_{\chi}$ for each population.  \SI{95}{\percent} confidence intervals from bootstrapping for each population are given by the error bars.  Colormap shows the probability distribution of $\sigma_{\chi}$ from bootstrapping.  Note distinct colorbars for each panel.  The \textit{clpX}E185$^*$ strain exhibits a statistically significant decline in $\sigma_{\langle \tau_r \rangle}$ and $\sigma_{\langle \tau_t \rangle}$ ($p<0.01$, permutation test), but not $\sigma_{\langle |v_r| \rangle}$. \label{fig_clpX}}
\end{centering}
\end{figure*}

Our abstract model suggests that such a reduction may arise from selection in the presence of a constraint on phenotypes.  From our numerical simulations, we show that the phenotype variation in a minimal model of directed evolution evolves as a function of the selection strength of population bottleneck and the number of selection rounds.  Within a broad range of parameter, the variance increases under extremely strong selection that always chooses the top $N_s$ largest phenotypes near the threshold where $N_s$ is the sample population at each selection round, while temperate selection allows accumulation of mutants with small isogenic fluctuations and hence can lead to decrease in the variance.  Thus our data suggest the possibility that swimming speed may be constrained in \textit{E. coli} by biophysical or metabolic means.  
%Our theoretical study suggests that these purely physical constraints may have important implications for the evolution of phenotypic fluctuations.  
Since there is no direct evidence for a threshold on traits such as run and tumble duration, the reduction in phenotypic fluctuations in run and tumble durations in our data could have a distinct mechanistic basis which may not be captured by our simple abstract model.  Another possible explanation for the reduced variance in run and tumble duration is that these traits evolve to lower values and are bounded by some lower bound, since cells cannot have infinitesimal run and tumble duration due to physical limitations.  This could make sense because in the soft agar gel the more frequently and the more quickly a cell switches its direction, the more efficiently it could find the correct gradient to do chemotaxis.  This is consistent with the experimental results shown in Fig. \ref{fig_experiment_variance}(A), where $P_p(\langle\tau_r\rangle)$ becomes more right-skewed as the mean of $P_p(\langle\tau_r\rangle)$ decreases over time, which is similar for $\langle\tau_t\rangle$.  Nevertheless, in Fig. \ref{fig_experiment_variance}(A) the left tail of $P_p(\langle\tau_r\rangle)$ does not clearly evolve towards the left even though $\overline{\langle\tau_r\rangle }$ decreases.  We suspect that the main phenotype subjected to the threshold could be a composite trait such as the run length which is the multiplication of run speed and run duration, and therefore the evolutionary trajectory of a single trait could become non-monotonous over time.  Further work is needed to elucidate the role of constraints on phenotypic fluctuations in run and tumble duration.

In addition, even though experimental data show a small increase or no significant decrease in variance between rounds \num{10} and \num{15} (Fig. \ref{fig_experiment_variance}(D-F)), the variance at round \num{15} is always less than variance of the founder and the increase is only significant for $\langle\tau_t\rangle$.  Therefore, this increase in variance from round \num{10} to \num{15} is at the limits of detectability and statistical significance in our experiments.  In our study of the abstract model as presented in Section \ref{sec_simulation_result}, when the mother-daughter correlation is high or the mutation range of the isogenic fluctuations ($\eta_{s_\chi}$) is small, phenotype variance can decrease even under very strong selection.  In these cases, if  mutation rate is high or the fluctuation range in threshold ($\eta_{\chi_c}$) is large, strong selection can still select mutant genotypes with large isogenic fluctuations ($s_{\chi}$) once mutants accumulate enough when $\overline\chi$ has evolved to near ${\chi_c}$, which can cause the variance in $P_p(\chi)$ to ``bounce back" and increase again as shown in Fig. S3.  If the model includes an intrinsic tendency to decline with increasing phenotype mean and the population amplification step (growth for $m$ generations) is not long enough to eliminate bias in the phenotype due to mother-daughter correlation, the bounce-back in variance could also appear due to the selection increasing the mean to a point whereby the intrinsic phenotype variance is smaller than the population phenotype variations.  Another logically-allowed possibility for the increase in variance is that the mutants that begin to dominate in the population at later rounds of selection have larger variance than the ones at earlier stages.  Finally, the specific constraint on the distributions due to the upper bound can also change the final variance.  However, these possibilities are parameter-dependent and thus are not necessary at the current stage especially since it is uncertain that the bounce-back is robust in the experiment. 
%
%In our study of the abstract model, if the model includes an intrinsic tendency to decline with increasing phenotype mean and population in liquid essay does not grow for enough generations to eliminate the selection bias on the phenotype, the bounce-back in variance could appear due to the selection increasing the mean to a point whereby the intrinsic phenotype variance is smaller than the population phenotype variations.  Another possibility that could cause similar effect is selecting individuals with phenotype within a fixed range of trait value, which is similar as selecting a fixed region of pipette size in the experiment, instead of selecting a fixed fraction of the population.  Under a fixed range of selection, a mutant with a sudden jump in its mean trait value and not too large isogenic fluctuations at intermediate rounds could lead to temporary large drop in variance, and once more mutants arise with large mean and various isogenic fluctuations at later rounds, the variance could rise again.   
%

Our abstract model of directed evolution applies to a broad range of potential systems and makes predictions of possible scenarios as to how the strength of selection can influence phenotypic fluctuations. 
%However, our abstract model of directed evolution applies to a broad range of potential systems and makes a clear prediction as to how the strength of selection influences phenotypic fluctuations. 
Genetic, biophysical and chemical constraints play an important role in the dynamics of biological systems from higher organisms such as fungi\cite{Roper:2008jj} to limits on the speed of protein translation \cite{Scott:2010cxa} and enzyme specificity\cite{Savir:2010hf}.  Our study highlights the potentially important role for these constraints in determining the limits of phenotypic fluctuations.  Future experimental evolution work could exploit known phenotypic constraints and directed evolution to directly test the predictions of our model.  

At a lower level of biological organization the mechanisms underlying phenotypic fluctuations remain hard to uncover in general due to the complex relationship between gene expression, protein function and cell-level phenotypes.  Despite the difficulty of connecting phenotypes to gene expression recent work has shown universal statistical properties in protein copy number distributions, with monotonically increasing scaling of the variance in protein abundances with mean expression levels \cite{taniguchi2010,Salman:2012bb,Brenner:2015gn,cohen2015}.  These universal properties of protein abundance fluctuations may provide a basis for understanding the evolution of phenotypic fluctuations in situations where the relevant regulatory architecture is known\cite{kaneko2009selection,Sato:2003uu}.   However, at present, a molecular accounting for the mechanism of the evolution of phenotype fluctuations requires detailed knowledge of the signaling pathways at work.  Our hope is that in studying abstract models such as the one presented here, we may uncover a more general understanding of when and why phenotypic fluctuations evolve.

%Our abstract simulations point to the importance of developing a coherent, abstracted, understanding of the relationship between selection, genetics and phenotypic fluctuations.  While quantitative geneticists have made important inroads in simple models of phenotypic evolution\cite{LANDE:1979th}, we propose that a quantitative understanding of the relationship between selection, fluctuations and the rate of evolution may be made through relatively simple scaling arguments similar to those from statistical physics.  

\ack
We thank Alvaro Hernandez at the Roy J Carver Biotechnology Center at the University of Illinois at Urbana-Champaign who performed the HiSeq sequencing and Elizabeth Ujhelyi who provided assistance with MiSeq sequencing.  We acknowledge partial support from National Science Foundation through grants PHY 0822613 and PHY 1430124 administered by the Center for Physics of Living Cells.  H.-Y. Shih was partially supported by the NASA Astrobiology Institute award NNA13AA91A, the L. S. Edelheit Family Fellowship in Biological Physics (2016) and the Government Scholarship for Study Abroad (2015-2016), Taiwan.
	
\clearpage

%%%%%%%%%% Selection procedure
%%%%%%%%%%

%%%%%  Phenotypic evolution %%%%%%%%%%%
%%%%%%%%%%%%%%%%%%%%%%

%%%%%%%%%%%%%%%%%%%%%%
%%%%%%%%%%%%%%%%%%%%%%

%%%%%%%%%% Selection procedure
%%%%%%%%%%

% \num{81} ($\Delta$1bp), \num{82} (\textit{clpX}) and \num{75} ($\Delta$1bp$+$\textit{clpX}) individuals

%\begin{figure}[ht!]
%\includegraphics[width=0.98\columnwidth]{abstract_thresold_random_variance.png}
%\caption{}\label{fig_abstract_simulation}
%\end{figure}

%\captionsetup[subfigure]{position=top, labelfont=bf,textfont=normalfont,singlelinecheck=off,justification=raggedright}
%\floatsetup[figure]{style=plain,subcapbesideposition=top}

%\captionsetup[subfigure]{position=top, labelfont=bf,textfont=normalfont,singlelinecheck=off,justification=raggedright}
%\floatsetup[figure]{style=plain,subcapbesideposition=top}

%%%%%%%%%% Selection procedure
%%%%%%%%%%

\clearpage
\bibliographystyle{ieeetr}
\bibliography{v4_chemotaxis_ref,v3_KuehnBib}

\begin{thebibliography}{10}

\bibitem{Elowitz:2002hb}
M.~B. Elowitz, A.~J. Levine, E.~D. Siggia, and P.~S. Swain, ``{Stochastic gene
  expression in a single cell.},'' {\em Science}, vol.~297, pp.~1183--1186,
  Aug. 2002.

\bibitem{Lehner:2010ei}
B.~Lehner, ``{Conflict between Noise and Plasticity in Yeast},'' {\em PLoS
  Genetics}, vol.~6, p.~e1001185, Nov. 2010.

\bibitem{Balaban:2004bq}
N.~Q. Balaban, J.~Merrin, R.~Chait, L.~Kowalik, and S.~Leibler, ``{Bacterial
  persistence as a phenotypic switch},'' {\em Science}, vol.~305,
  pp.~1622--1625, Sept. 2004.

\bibitem{Spudich:1976wv}
J.~Spudich and D.~Koshland, ``{Non-genetic individuality: chance in the single
  cell},'' {\em Nature}, vol.~262, pp.~467--471, Aug. 1976.

\bibitem{Frankel:2014fb}
N.~W. Frankel, W.~Pontius, Y.~S. Dufour, J.~Long, L.~Hernandez-Nunez, and
  T.~Emonet, ``{Adaptability of non-genetic diversity in bacterial
  chemotaxis},'' {\em eLife}, vol.~3, p.~e03526, Oct. 2014.

\bibitem{Baldwin:1896dt}
J.~M. Baldwin, ``{A new factor in evolution},'' {\em The American Naturalist},
  1896.

\bibitem{Waddington:1953vf}
C.~H. Waddington, ``{Genetic assimilation of an acquired character},'' {\em
  Evolution}, vol.~7, pp.~118--126, Oct. 1953.

\bibitem{WestEberhard:1989wq}
M.~West-Eberhard, ``{Phenotypic plasticity and the origins of diversity},''
  {\em Annual Review of Ecology, Evolution and Systematics}, vol.~20,
  pp.~249--278, Nov. 1989.

\bibitem{Sato:2003uu}
K.~Sato, Y.~Ito, T.~Yomo, and K.~Kaneko, ``{On the relation between fluctuation
  and response in biological systems},'' {\em Proceedings of the National
  Academy of Sciences of USA}, vol.~100, p.~14086, Nov. 2003.

\bibitem{kaneko2009selection}
Y.~Ito, H.~Toyota, K.~Kaneko, and T.~Yomo, ``How selection affects phenotypic
  fluctuation.,'' {\em Molecular systems biology}, vol.~5, pp.~264--264, Jan.
  2009.

\bibitem{yomo2014}
M.~Yoshida, S.~Tsuru, N.~Hirata, S.~Seno, H.~Matsuda, B.-W. Ying, and T.~Yomo,
  ``Directed evolution of cell size in escherichia coli,'' {\em BMC
  Evolutionary Biology}, vol.~14, pp.~257--257, Dec. 2014.

\bibitem{kaneko2005}
C.~Furusawa, T.~Suzuki, A.~Kashiwagi, T.~Yomo, and K.~Kaneko, ``Ubiquity of
  log-normal distributions in intra-cellular reaction dynamics,'' {\em
  Biophysics}, vol.~1, pp.~25--31, Apr. 2005.

\bibitem{Salman:2012bb}
H.~Salman, N.~Brenner, C.-k. Tung, N.~Elyahu, E.~Stolovicki, L.~Moore,
  A.~Libchaber, and E.~Braun, ``{Universal Protein Fluctuations in Populations
  of Microorganisms},'' {\em Physical Review Letters}, vol.~108, p.~238105,
  June 2012.

\bibitem{cohen2015}
M.~S. Sherman, K.~Lorenz, M.~H. Lanier, and B.~A. Cohen, ``Cell-to-cell
  variability in the propensity to transcribe explains correlated fluctuations
  in gene expression,'' {\em Cell systems}, vol.~1, pp.~315--325, Nov. 2015.

\bibitem{taniguchi2010}
Y.~Taniguchi, P.~J. Choi, G.-W. Li, H.~Chen, M.~Babu, J.~Hearn, A.~Emili, and
  X.~S. Xie, ``Quantifying e. coli proteome and transcriptome with
  single-molecule sensitivity in single cells,'' {\em science}, vol.~329,
  pp.~533--538, July 2010.

\bibitem{salman2015single}
N.~Brenner, E.~Braun, A.~Yoney, L.~Susman, J.~Rotella, and H.~Salman,
  ``Single-cell protein dynamics reproduce universal fluctuations in cell
  populations,'' {\em The European Physical Journal E}, vol.~38, p.~102, Sept.
  2015.

\bibitem{salman2016}
L.~Susman, M.~Kohram, H.~Vashistha, J.~T. Nechleba, H.~Salman, and N.~Brenner,
  ``Statistical properties and dynamics of phenotype components in individual
  bacteria,'' {\em arXiv preprint arXiv:1609.05513}, 2016.

\bibitem{Milo:2007wm}
R.~Milo, J.~H. Hou, M.~Springer, M.~P. Brenner, and M.~W. Kirschner, ``{The
  relationship between evolutionary and physiological variation in
  hemoglobin},'' {\em Proceedings of the National Academy of Sciences of USA},
  vol.~104, p.~16998, Oct. 2007.

\bibitem{Barrick:2009in}
J.~E. Barrick, D.~S. Yu, S.~H. Yoon, H.~Jeong, T.~K. Oh, D.~Schneider, R.~E.
  Lenski, and J.~F. Kim, ``{Genome evolution and adaptation in a long-term
  experiment with Escherichia coli.},'' {\em Nature}, vol.~461, pp.~1243--1247,
  Oct. 2009.

\bibitem{Rainey:1998fx}
P.~B. Rainey and M.~Travisano, ``{Adaptive radiation in a heterogeneous
  environment},'' {\em Nature}, vol.~394, pp.~69--72, July 1998.

\bibitem{Bachmann:2013dm}
H.~Bachmann, M.~Fischlechner, I.~Rabbers, N.~Barfa, F.~Branco~dos Santos,
  D.~Molenaar, and B.~Teusink, ``{Availability of public goods shapes the
  evolution of competing metabolic strategies.},'' {\em Proceedings of the
  National Academy of Sciences of USA}, vol.~110, pp.~14302--14307, Aug. 2013.

\bibitem{Fraebel:2017wx}
D.~T. Fraebel, H.~Mickalide, D.~Schnitkey, J.~Merritt, T.~E. Kuhlman, and
  S.~Kuehn, ``{Environment determines evolutionary trajectory in a constrained
  phenotypic space},'' {\em eLife}, vol.~6, p.~e24669, Mar. 2017.

\bibitem{Wolfe:1989ue}
A.~J. Wolfe and H.~C. Berg, ``{Migration of Bacteria in Semisolid Agar},'' {\em
  Proceedings of the National Academy of Sciences of USA}, vol.~86,
  pp.~6973--6977, Sept. 1989.

\bibitem{Adler:1966wi}
J.~Adler, ``{Chemotaxis in Bacteria},'' {\em Science}, vol.~153, pp.~708--716,
  Aug. 1966.

\bibitem{Yi:2016dp}
X.~Yi and A.~M. Dean, ``{Phenotypic plasticity as an adaptation to a functional
  trade-off},'' {\em eLife}, vol.~5, pp.~1--12, Oct. 2016.

\bibitem{Korobkova:2004vs}
E.~Korobkova, T.~Emonet, J.~M. Vilar, T.~Shimizu, and P.~Cluzel, ``{From
  molecular noise to behavioural variability in a single bacterium},'' {\em
  Nature}, vol.~428, pp.~574--578, Apr. 2004.

\bibitem{Bai:2013eea}
F.~Bai, Y.-S. Che, N.~Kami-ike, Q.~Ma, T.~Minamino, Y.~Sowa, and K.~Namba,
  ``{Populational Heterogeneity vs. Temporal Fluctuation in Escherichia coli
  Flagellar Motor Switching},'' {\em Biophysj}, vol.~105, pp.~2123--2129, Nov.
  2013.

\bibitem{Jordan:2013hf}
D.~Jordan, S.~Kuehn, E.~Katifori, and S.~Leibler, ``{Behavioral diversity in
  microbes and low-dimensional phenotypic spaces},'' {\em Proceedings of the
  National Academy of Sciences of USA}, vol.~110, pp.~14018--14023, May 2013.

\bibitem{wagner1996}
G.~P. Wagner and L.~Altenberg, ``Perspective: complex adaptations and the
  evolution of evolvability,'' {\em Evolution}, vol.~50, pp.~967--976, June
  1996.

\bibitem{lande1976}
R.~Lande, ``Natural selection and random genetic drift in phenotypic
  evolution,'' {\em Evolution}, vol.~30, no.~2, pp.~314--334, 1976.

\bibitem{jensen2000}
J.~L. Jensen, {\em Statistics for petroleum engineers and geoscientists},
  vol.~2.
\newblock Gulf Professional Publishing, 2000.

\bibitem{raser2005}
J.~M. Raser and E.~K. O'shea, ``Noise in gene expression: origins,
  consequences, and control,'' {\em Science}, vol.~309, pp.~2010--2013, Sept.
  2005.

\bibitem{feinberg2010}
A.~P. Feinberg and R.~A. Irizarry, ``Stochastic epigenetic variation as a
  driving force of development, evolutionary adaptation, and disease,'' {\em
  Proceedings of the National Academy of Sciences}, vol.~107, pp.~1757--1764,
  Jan. 2010.

\bibitem{hill2004}
W.~G. Hill and X.-s. Zhang, ``Effects on phenotypic variability of directional
  selection arising through genetic differences in residual variability,'' {\em
  Genetics Research}, vol.~83, pp.~121--132, May 2004.

\bibitem{elowitz2010}
A.~Eldar and M.~B. Elowitz, ``Functional roles for noise in genetic circuits,''
  {\em Nature}, vol.~467, pp.~167--173, Sept. 2010.

\bibitem{waddington1942}
C.~H. Waddington, ``Canalization of development and the inheritance of acquired
  characters,'' {\em Nature}, vol.~150, pp.~563--565, Nov. 1942.

\bibitem{lindquist1998}
S.~L. Rutherford and S.~Lindquist, ``Hsp90 as a capacitor for morphological
  evolution,'' {\em Nature}, vol.~396, pp.~336--342, Nov. 1998.

\bibitem{Darnton:2007cta}
N.~C. Darnton, L.~Turner, S.~Rojevsky, and H.~C. Berg, ``{On Torque and
  Tumbling in Swimming Escherichia coli},'' {\em Journal of Bacteriology},
  vol.~189, pp.~1756--1764, Feb. 2007.

\bibitem{Boehm:2010du}
A.~Boehm, M.~Kaiser, H.~Li, C.~Spangler, C.~A. Kasper, M.~Ackermann, V.~Kaever,
  V.~Sourjik, V.~Roth, and U.~Jenal, ``{Second Messenger-Mediated Adjustment of
  Bacterial Swimming Velocity},'' {\em Cell}, vol.~141, pp.~107--116, Apr.
  2010.

\bibitem{Dufour:2016bo}
Y.~S. Dufour, S.~Gillet, N.~W. Frankel, D.~B. Weibel, and T.~Emonet, ``{Direct
  Correlation between Motile Behavior and Protein Abundance in Single Cells},''
  {\em PLoS Computational Biology}, vol.~12, pp.~e1005041--25, Sept. 2016.

\bibitem{Vladimirov:2008ik}
N.~Vladimirov, L.~L{\o}vdok, D.~Lebiedz, and V.~Sourjik, ``{Dependence of
  Bacterial Chemotaxis on Gradient Shape and Adaptation Rate},'' {\em PLoS
  Computational Biology}, vol.~4, pp.~--e1000242, Dec. 2008.

\bibitem{Kalir:2001gr}
S.~Kalir, J.~McClure, K.~Pabbaraju, C.~Southward, M.~Ronen, S.~Leibler,
  M.~Surette, and U.~Alon, ``{Ordering genes in a flagella pathway by analysis
  of expression kinetics from living bacteria},'' {\em Science}, vol.~292,
  p.~2080, June 2001.

\bibitem{Tomoyasu:2003dx}
T.~Tomoyasu, A.~Takaya, E.~Isogai, and T.~Yamamoto, ``{Turnover of FlhD and
  FlhC, master regulator proteins for Salmonella flagellum biogenesis, by the
  ATP-dependent ClpXP protease.},'' {\em Molecular microbiology}, vol.~48,
  pp.~443--452, Apr. 2003.

\bibitem{Zhao:2007dz}
K.~Zhao, M.~Liu, and R.~R. Burgess, ``{Adaptation in bacterial flagellar and
  motility systems: from regulon members to
  {\textquoteleft}foraging{\textquoteright}-like behavior in E. coli},'' {\em
  Nucleic Acids Research}, vol.~35, pp.~4441--4452, June 2007.

\bibitem{Lovdok:2008jj}
L.~E. L{\o}vdok, {\em {Gene expression noise and robustness in the Escherichia
  coli chemotaxis pathway}}.
\newblock PhD thesis, University of Heidelberg, 2008.

\bibitem{Roper:2008jj}
M.~Roper, R.~E. Pepper, M.~P. Brenner, and A.~Pringle, ``{Explosively launched
  spores of ascomycete fungi have drag-minimizing shapes.},'' {\em Proceedings
  of the National Academy of Sciences of USA}, vol.~105, pp.~20583--20588, Dec.
  2008.

\bibitem{Scott:2010cxa}
M.~Scott, C.~W. Gunderson, E.~M. Mateescu, Z.~Zhang, and T.~Hwa,
  ``{Interdependence of cell growth and gene expression: origins and
  consequences.},'' {\em Science}, vol.~330, pp.~1099--1102, Nov. 2010.

\bibitem{Savir:2010hf}
Y.~Savir, E.~Noor, R.~Milo, and T.~Tlusty, ``{Cross-species analysis traces
  adaptation of Rubisco toward optimality in a low-dimensional landscape},''
  {\em Proceedings of the National Academy of Sciences of USA}, vol.~107,
  pp.~3475--3480, Feb. 2010.

\bibitem{Brenner:2015gn}
N.~Brenner, C.~M. Newman, D.~Osmanovi{\'c}, Y.~Rabin, H.~Salman, and D.~L.
  Stein, ``{Universal protein distributions in a model of cell growth and
  division},'' {\em Physical Review E}, vol.~92, pp.~042713--6, Oct. 2015.

\end{thebibliography}

\end{document}